\ifdefined\pdfminorversion\pdfminorversion=7\relax\fi
\PassOptionsToPackage{pdfversion=1.7}{hyperref}

\ifdefined\pdfsuppressptexinfo\pdfsuppressptexinfo=8\relax\fi

\documentclass[sigconf,screen,nonacm]{acmart}

\setcopyright{none}
\AtBeginDocument{\renewcommand{\balance}{}}
\AtBeginDocument{}

\usepackage{amsmath,amssymb,amsfonts}
\usepackage{algorithm}
\usepackage{algpseudocode}
\usepackage{graphicx}
\usepackage{xcolor}
\usepackage{listings}
\usepackage{booktabs}
\usepackage{multirow}
\usepackage{pifont}
\usepackage{subcaption}
\usepackage{balance}

\newcounter{obs}

\graphicspath{{images/}}

\usepackage{xspace}
\newcommand{\sysnametext}{FlatCake}
\newcommand{\sys}{\textup{\texttt{\sysnametext}}\xspace}
\newcommand{\sysn}{\sysnametext\xspace}
\newcommand{\modpfx}{FC}
\newcommand{\gva}{\modpfx-GVA\xspace}
\newcommand{\mas}{\modpfx-MAS\xspace}
\newcommand{\cio}{\modpfx-CIO\xspace}

\newcommand{\rih}[1]{\noindent\textbf{#1.}}

\usepackage[normalem]{ulem}

\usepackage{microtype}         
\newcommand{\authsep}{\quad}
\newcommand{\namesep}{\hspace{1.5em}}
\newcommand{\authcomma}{,}
\begin{document}

\newcommand{\papertitle}{A Flat KV Cache Storage Architecture for LLM Serving}
\title[\papertitle]{\papertitle}


\author{%
  \large\rmfamily
  \begin{tabular}[t]{@{}c@{}}
    {\LARGE\rmfamily
      Xin Huang\textsuperscript{1\authcomma2}\namesep
      Teng Ma\textsuperscript{3}\namesep
      Zhuoran Li\textsuperscript{4}\namesep
      Yibo Zhao\textsuperscript{4}\namesep
      Zhaoshi Li\textsuperscript{5}\namesep
      Qianchao Zhu\textsuperscript{3}}\\[2pt]
    {\LARGE\rmfamily
      Wei Yang\textsuperscript{1\authcomma2}\namesep
      Chenxi Wang\textsuperscript{6\authcomma2}\namesep
      Tao Xie\textsuperscript{4\authcomma1\authcomma2\authcomma*}\namesep
      Youwei Zhuo\textsuperscript{4\authcomma2\authcomma*}}\\[5pt]
    \textsuperscript{1}Fudan University\authcomma{} China\authsep
      \textsuperscript{2}Beijing Tongming Lake Center\authcomma{} China\authsep
      \textsuperscript{3}Independent Researcher\\
    \textsuperscript{4}Peking University\authcomma{} China\authsep
      \textsuperscript{5}MetaX Technology\authcomma{} China\authsep
      \textsuperscript{6}Institute of Computing Technology\authcomma{} CAS\authcomma{} China\\[3pt]
    huangxin25@m.fudan.edu.cn\authsep
      stmatengss@gmail.com\authsep
      \{zhuoranli\authcomma{} yibozhao\}@stu.pku.edu.cn\\
    zhaoshi.li@metax-tech.com\authsep
      wangbo18@xiaohongshu.com\authsep
      yang\_wei@fudan.edu.cn\\
    wangchenxi@ict.ac.cn\authsep
      \{taoxie\authcomma{} youwei\}@pku.edu.cn\authsep
      \textsuperscript{*}Corresponding authors
  \end{tabular}%
}
\renewcommand{\authors}{Xin Huang\and Teng Ma\and Zhuoran Li\and Yibo Zhao\and
Zhaoshi Li\and Qianchao Zhu\and Wei Yang\and Chenxi Wang\and Tao Xie\and Youwei Zhuo}

\renewcommand{\shortauthors}{Xin Huang, Teng Ma, Zhuoran Li, Yibo Zhao, Zhaoshi Li,
Qianchao Zhu, Wei Yang, Chenxi Wang, Tao Xie, and Youwei Zhuo}



\begin{abstract}
KV cache reuse accelerates LLM serving,
but existing systems manage host DRAM and NVMe SSD as a tiered hierarchy
where DRAM acts as an intermediate cache.
Production LLM workloads break both premises of this hierarchy.
First, bimodal reuse defeats DRAM temporal filtering,
since workloads are dominated either by one-shot blocks with long reuse distances
or by multi-gigabyte agentic sessions that exceed DRAM capacity,
putting SSD on the critical read path.
Second, DRAM staging and eviction-ordered layout bottleneck SSD restoration.
Holding host staging buffers for in-flight prefixes causes contention under load
that aborts prefetches, while incremental eviction scatters a session's blocks
across the SSD and prevents large sequential reads.

We present \sys, a \emph{flat} KV storage architecture that exposes host DRAM and NVMe SSD
as peer media in a unified 64-bit address space.
Exploiting the fact that KV blocks are write-once, immutable, and reused together within a prompt or session,
\sys places each block directly onto its target medium in session-contiguous, layer-major order
at creation time and never migrates it.
\sys streams coalesced SSD reads directly into a bounded GPU window via GPUDirect Storage
and decouples DRAM and SSD request queues to prevent head-of-line blocking.
We implement \sys in SGLang and compare it against host-staged~(Mooncake) and GDS-enabled~(Mooncake-GDS) tiered storage,
as well as SSD-direct storage~(Tutti).
Across three production agentic, coding, and conversational traces with DeepSeek-V4-Flash,
\sys reduces mean TTFT near the baselines' latency knees by $1.40\times$--$2.42\times$ over the fastest baseline.
On the agentic Claude Code trace, \sys sustains a 91.6\% cache hit rate under load~(versus 71.1\% for all three baselines)
and delivers up to $2.16\times$ higher token throughput.
\end{abstract}

\maketitle

\section{Introduction}
\label{sec:intro}

Prefix caching is now standard in large language model~(LLM) serving,
allowing requests that share a prefix to reuse previously computed key-value~(KV) tensors
and skip redundant prefill computation~\cite{vllm2023,sglang2024,mooncake2024}.
Early workloads mainly reused short, globally shared system prompts~\cite{kvcache_wild2026},
whereas modern serving mixes multi-turn conversations~\cite{bidaw2026} with long-context agentic coding sessions
that append tool outputs and retrieved files on every turn.
In a production Claude Code trace, requests average 219K input tokens~\cite{ccTracesWeka2026},
pushing active KV working sets to tens of gigabytes per session over multiple hours.
Because concurrent working sets exceed GPU HBM and host DRAM capacity
and recomputing evicted history consumes up to 93\% of prefill compute~\cite{bidaw2026},
NVMe SSDs must serve KV cache hits on the critical path~\cite{tutti2025}.

\begin{figure}[!t]
  \centering
  \includegraphics[width=\columnwidth]{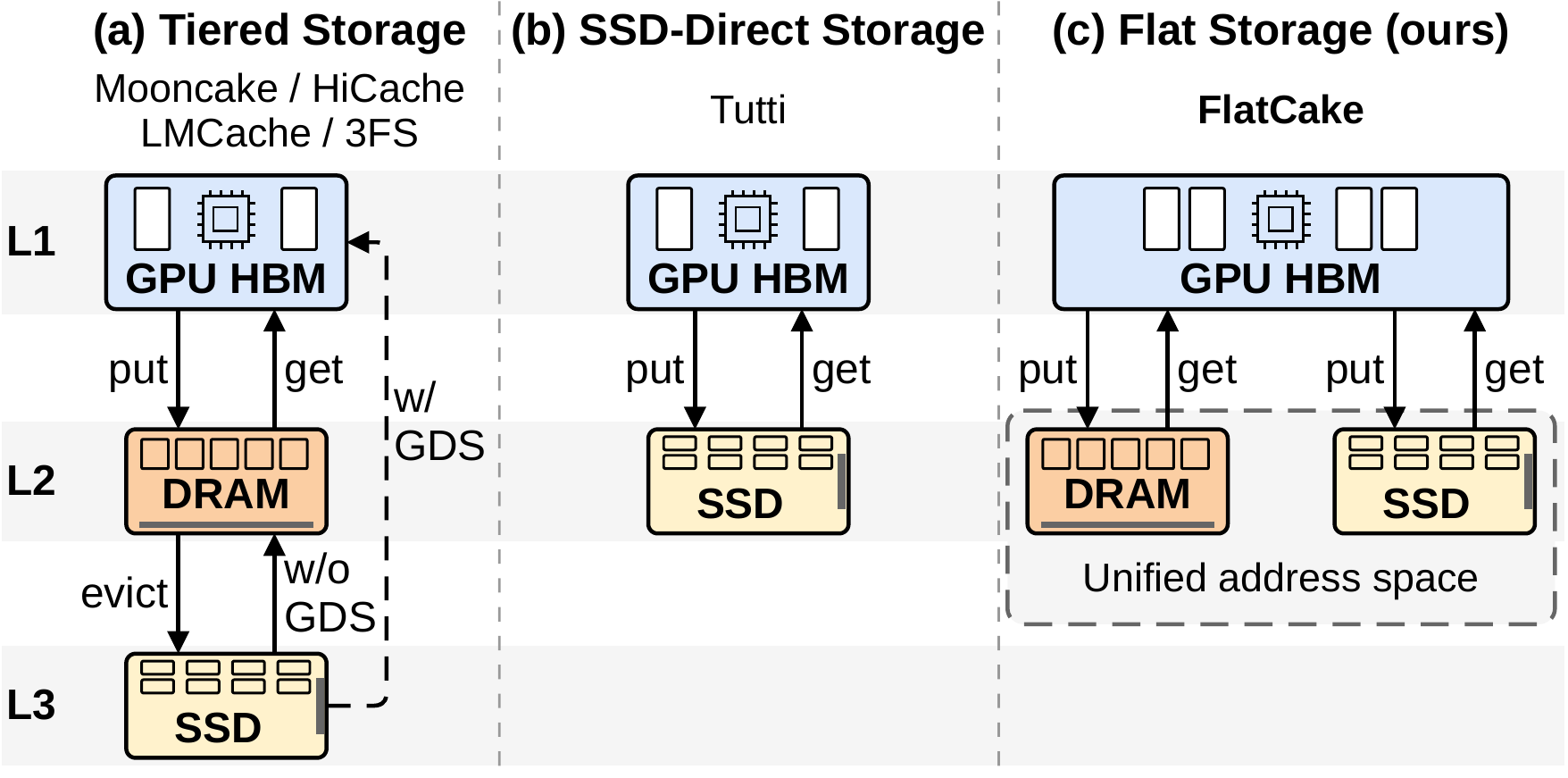}
  \caption{KV cache storage organizations.
  (a)~Tiered storage
  admits blocks into host DRAM first, evicts cold blocks to SSD,
  and restores SSD hits through host DRAM~(w/o GDS) or a direct GDS bypass~(w/ GDS).
  (b)~SSD-direct storage
  transfers KV blocks directly between GPU HBM and SSD
  via GPU-executed I/O, leaving host DRAM unused.
  (c)~Flat storage~(\sys) exposes DRAM and SSD as peers in a unified address space
  and writes each block straight to its target medium without cross-media migration.}
  \Description{Three KV cache storage organizations side by side, drawn as
  hardware icons on shaded L1, L2 and L3 rows, with a GPU HBM icon at L1 in every
  panel. In panel (a), Tiered Storage, DRAM is at L2 and SSD at L3: put and get
  arrows join HBM and DRAM, an evict arrow goes from DRAM to SSD, a w/o GDS arrow
  returns from SSD to DRAM, and a dashed w/ GDS arrow goes from SSD directly to
  HBM. Panel (b), SSD-Direct Storage, shows Tutti with put and get arrows between
  HBM and an L2 SSD, without DRAM. Panel (c), Flat Storage, shows \sysnametext{}
  with DRAM and SSD side by side at L2 inside a dashed unified address space;
  each medium has its own put and get arrows to HBM, and there is no arrow
  between the two media. Light vertical dashed lines separate the three panels.}
  \label{fig:arch_comparison}
\end{figure}

Existing systems such as Mooncake~\cite{mooncake2024}, LMCache~\cite{lmcache2025},
and SGLang HiCache~\cite{sglang_hicache2025} organize GPU HBM, host DRAM, and NVMe SSD
as a \emph{tiered hierarchy}~(Figure~\ref{fig:arch_comparison}(a)).
By admitting newly generated KV blocks into host DRAM first and evicting cold blocks to SSD in the background,
tiering absorbs prefill writes at memory bandwidth and serves recent hits without touching slower block storage.
Borrowed from general-purpose memory and storage systems,
this design rests on two premises.
First, \emph{temporal filtering} assumes that recency-based DRAM caching retains the active working set
so NVMe SSD serves only cold misses off the critical path.
Second, \emph{eviction-driven offloading} assumes that populating SSD incrementally through DRAM eviction
and staging restores through host DRAM buffers can sustain high read throughput whenever SSD hits occur.
Characterizing eight production traces from Mooncake, Qwen Bailian, and Claude Code~(\S\ref{sec:motivation})
shows that LLM serving workloads break both premises.

First, bimodal reuse defeats recency-based DRAM filtering
and puts NVMe SSD on the critical read path.
Across the eight traces, KV reuse splits into two regimes.
In conversational and API workloads, 53--94\% of KV blocks are accessed only once,
whereas reusable prefixes span hundreds to thousands of intervening requests.
Admitting every write into host DRAM pollutes it with one-shot blocks
that evict compact, reusable prefixes to SSD before their next turn arrives.
Conversely, in agentic coding~(Claude Code), 83.4\% of blocks are reused with immediate turn-to-turn locality,
yet multi-gigabyte session histories quickly exceed host DRAM capacity across concurrent sessions.
In both regimes, funneling all writes through an LRU DRAM tier fails to keep hot prefixes in memory
and pushes most cached token volume onto NVMe SSD,
turning DRAM--SSD migration into wasted data movement.

Second, host DRAM staging and eviction-ordered layout turn SSD restoration
into a concurrency and I/O bottleneck.
Once SSD must serve the bulk of cached tokens, both tiered read paths in Figure~\ref{fig:arch_comparison}(a) degrade under load.
On the host-staged path, restoring a long prefix requires holding host DRAM buffers for the entire in-flight hit until GPU transfers finish,
so staging memory scales with context length and concurrency.
Even when SSD-to-DRAM and DRAM-to-GPU copies are pipelined, concurrent restores contend for bounded staging buffers,
delaying or aborting prefetches and forcing the GPU to recompute KV blocks already stored on SSD.
Bypassing host DRAM with batched GPUDirect Storage~(GDS) avoids staging contention
but still reads an eviction-scrambled disk layout.
Because tiered systems write blocks to SSD in DRAM eviction order
and pack evicted blocks from different sessions into shared fixed-size FIFO buckets,
each session's prefix is scattered across non-contiguous SSD offsets.
Restoring a prefix layer by layer therefore issues $O(\text{layers} \times \text{blocks})$ small reads that underutilize sequential NVMe bandwidth,
while recycling an old bucket invalidates active prefixes across sessions.

SSD-direct systems such as Tutti~\cite{tutti2025} avoid cross-tier migration and host staging contention
by transferring KV blocks directly between GPU HBM and NVMe SSD
via GPU-executed I/O queues~(Figure~\ref{fig:arch_comparison}(b)).
However, bypassing host DRAM leaves hundreds of gigabytes to terabytes of server memory unused,
even though it delivers $2.4\text{--}2.5\times$ higher read bandwidth than local NVMe SSDs~(Table~\ref{tab:hardware})
and can serve compact shared prompts and short initial turns without occupying GPU threads for I/O.
Furthermore, Tutti packs all layers of each KV page together~(page-major layout),
so whole-page SSD reads on long-context sessions complete layer~0 only after the last page arrives
and cannot overlap with layer-wise computation.
Neither placing DRAM as a migratory filter above SSD nor discarding DRAM altogether
matches how LLM KV caches are used.

Why do existing designs face this trade-off?
General-purpose tiering assumes mutable pages whose future reuse and spatial grouping are unknown when written,
relying on dynamic migration through DRAM to discover locality.
Our key insight is that LLM KV cache blocks are \emph{write-once, immutable,
and reused together as a shared prompt or session prefix}.
Because blocks are never modified in place, they carry no dirty state to write back.
Moreover, both a request's prefix sharing and its write size are observable at creation time,
and every subsequent turn restores a session's blocks together layer by layer.
Compact shared prompts and short writes can therefore stay pinned in fast host DRAM without one-shot pollution,
while large writes and multi-gigabyte sessions can be written straight to NVMe SSD in session-contiguous, layer-major order
so restores stream sequentially and overlap computation without DRAM migration or reassembly.

Based on this insight, we propose \sys, a \emph{flat KV storage architecture}
that unifies host DRAM and NVMe SSD as peer storage media~(\S\ref{sec:design}).
As shown in Figure~\ref{fig:arch_comparison}(c), \sys exposes both media at the same level
behind a unified 64-bit Global Virtual Address~(GVA) space,
placing each KV block once at write time without cross-media migration.
Managing peer media without an intermediate DRAM buffer raises challenges in initial placement,
cross-medium scheduling, and direct SSD streaming, which \sys addresses through three components:
\begin{itemize}
\item \textbf{\gva: Global Virtual Address Manager}~(\S\ref{sec:gva}).
\gva binds each newly generated prefix or session~(a \emph{reuse scope}) directly to DRAM or SSD at creation
using observable sharing and write-size signals,
where any SSD-placed scope still reads sequentially rather than triggering migration.
Encoding each block's medium, node, and offset in an immutable 64-bit GVA enables $O(1)$ media-aware routing
and zero-writeback, whole-session SSD reclamation.

\item \textbf{\mas: Medium-Aware Request Scheduler}~(\S\ref{sec:scheduler}).
Without DRAM pre-staging, slow SSD restores can stall sub-millisecond DRAM hits in a shared queue,
whereas strict DRAM priority can starve large SSD sessions.
Using the medium bits in each GVA, \mas separates DRAM-ready and SSD-preparing queues,
prioritizes smaller SSD prefetches while aging waiting ones,
and re-admits completed prefetches in original arrival order to prevent both head-of-line blocking and starvation.

\item \textbf{\cio: Coalescing Direct I/O Engine}~(\S\ref{sec:gds}).
\cio combines write-time layer-major layout within pre-allocated session files and read-time descriptor coalescing
to collapse $O(\text{layers} \times \text{blocks})$ fragmented reads into $O(\text{layers})$ sequential GDS transfers.
Because paged KV slots in GPU HBM are non-contiguous, \cio streams each layer extent into a bounded GPU staging window
overlapped with layer-wise computation before scattering pages into KV slots, eliminating host staging contention.
\end{itemize}

We implement \sys in SGLang and compare it against host-staged~(Mooncake)
and GDS-enabled~(Mooncake-GDS) tiered storage, as well as SSD-direct storage~(Tutti),
across production trace replay and long-context workloads~(\S\ref{sec:eval}).
Replaying three production traces~(Claude Code, Qwen Bailian Coder, and Mooncake Conversation)
on DeepSeek-V4-Flash with NVIDIA B200 GPUs,
\sys reduces mean TTFT near the baselines' latency knees by $1.40\times$--$2.42\times$ over the fastest baseline.
On Claude Code, where all three SSD-backed baselines drop to a 71.1\% delivered cache hit rate under load,
\sys sustains 91.6\% and delivers up to $2.16\times$ higher token throughput.

\section{Background and Motivation}
\label{sec:background}

\subsection{KV Cache Reuse and Tiered Storage}
\label{sec:reuse_and_tiering}
Once generated during prefill, KV cache~\cite{attention2017} blocks are write-once and immutable,
and their reuse across requests follows three primary patterns
determined by how the shared prefix is formed.
\emph{Prefix caching}~\cite{sglang2024} reuses KV blocks
\emph{across} requests that share a static prefix, such as a system prompt or few-shot template,
so only the uncached suffix requires prefill computation.
\emph{Multi-turn conversations} reuse KV blocks \emph{within} a user session,
where each turn appends a new prompt and model response to the accumulated history of prior turns.
\emph{Agentic workloads} also accumulate state within a session,
but append tool invocations, retrieved files, and command outputs on every turn
while spawning sub-agents that replay the shared session prefix in parallel.
As a result, per-turn context grows by tens of thousands of tokens,
and a single session's KV footprint can reach tens of gigabytes.

To accommodate working sets that exceed GPU HBM, existing systems such as
Mooncake~\cite{mooncake2024}, LMCache~\cite{lmcache2025}, and SGLang HiCache~\cite{sglang_hicache2025}
organize GPU HBM, host DRAM, and NVMe SSD as a three-level \emph{tiered hierarchy}~(Figure~\ref{fig:arch_comparison}(a)).
Newly generated KV blocks enter host DRAM first to absorb prefill writes at memory speed,
spill incrementally to NVMe SSD under LRU or FIFO eviction when DRAM fills,
and return to the GPU either through host DRAM staging buffers or via a direct GPUDirect Storage~(GDS) bypass.
This hierarchy rests on two premises.
First, recency-based \emph{temporal filtering} in host DRAM keeps the active working set in memory
so NVMe SSD serves only cold misses off the critical path.
Second, \emph{eviction-driven offloading} and host-staged or direct restoration
sustain high read throughput whenever SSD hits occur.

\begin{table}[t]
  \centering
  \caption{KV cache reuse statistics across eight production LLM traces under an infinite-capacity cache.}
  \label{tab:hitrate}
  \footnotesize
  \setlength{\tabcolsep}{0pt}
  \begin{tabular*}{\columnwidth}{@{\extracolsep{\fill}}lrrrrrc@{}}
    \toprule
    & & Unique & Once- & \multicolumn{2}{c}{Reuse dist} & Hit \\
    \cmidrule(lr){5-6}
    Workload & Requests & blocks & access & Mean & P99 & rate \\
    \midrule
    \multicolumn{7}{@{}l}{\textit{Mooncake Traces}~\cite{mooncake2024}} \\
    \quad Conversation & 12,031  & 182,790   & 75.8\% & 728 & 5,527  & 36.64\% \\
    \quad Synthetic    & 3,993   & 43,924    & 58.4\% & 430 & 2,235  & 63.96\% \\
    \quad ToolAgent    & 23,608  & 183,300   & 78.5\% & 489 & 4,085  & 55.26\% \\
    \midrule
    \multicolumn{7}{@{}l}{\textit{Qwen Bailian Traces}~\cite{kvcache_wild2026}} \\
    \quad to-C         & 43,058  & 2,530,337 & 53.2\% & 1,450 & 13,973 & 59.4\% \\
    \quad to-B API     & 172,800 & 3,830,390 & 94.0\% & 2,550 & 54,728 & 61.3\% \\
    \quad Thinking     & 10,812  & 1,735,015 & 85.7\% & 155  & 2,357  & 46.2\% \\
    \quad Coder        & 43,011  & 5,202,517 & 61.0\% & 1,095 & 16,224 & 66.4\% \\
    \midrule
    \multicolumn{7}{@{}l}{\textit{SemiAnalysis Agentic Trace}~\cite{ccTracesWeka2026}} \\
    \quad Claude Code  & 98,827  & 5,717,185 & 16.6\% & 0.85  & 14     & 98.31\% \\
    \bottomrule
  \end{tabular*}
\end{table}

\subsection{Why Tiered Storage Fails on LLM Workloads}
\label{sec:motivation}
\label{sec:flat_motivation}
We evaluate both premises across eight production traces from Mooncake, Qwen Bailian,
and SemiAnalysis.
Table~\ref{tab:hitrate} also reports the mean and P99 \emph{reuse distance} across reuse events,
defined as the number of intervening requests between consecutive accesses to the same block
.

\refstepcounter{obs}\label{obs:hitrate}%
\noindent\textbf{Observation~\theobs: Bimodal reuse defeats DRAM temporal filtering and puts SSD on the critical read path.}
Infinite-capacity hit rates span 36.6\% to 98.3\% across the eight traces,
yet the workloads split into two reuse regimes that both defeat recency-based DRAM caching.
Because an LRU or FIFO tier admits every newly generated block into DRAM,
it retains a reused block only if all distinct blocks written across the intervening requests fit within DRAM capacity.

Across the seven prefix-caching and conversation traces, 53.2--94.0\% of blocks
are accessed only once, peaking at 94.0\% in Qwen to-B API and 78.5\% in Mooncake ToolAgent.
While the top 10\% of blocks account for 34--65\% of all accesses,
reuse distances exhibit heavy tails, reaching a mean of 2{,}550 intervening requests
and a P99 of 54{,}728 in Qwen to-B API~(489 and 4{,}085 in Mooncake ToolAgent).
When every newly generated block is admitted into host DRAM first,
one-shot blocks continually flush reusable session prefixes to SSD before their next turn arrives.
Replacing the DRAM eviction policy yields little benefit,
as LRU, FIFO, and queue-enhanced policies all achieve DRAM hit rates near 20\%
even when the fast tier holds 40.1\% of total blocks in multi-turn serving~\cite{bidaw2026}.

\begin{figure}[t]
  \centering
  \includegraphics[width=0.9\columnwidth]{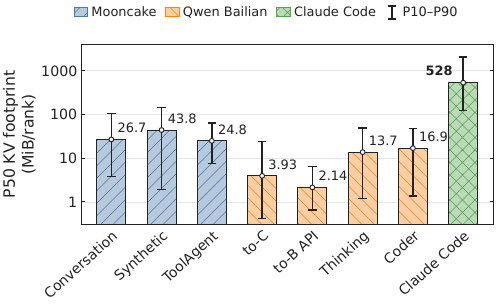}
  \caption{Per-request FP8 KV cache footprint per GPU rank on DeepSeek-V4-Flash~(TP8),
  derived from trace block counts across the eight workloads,
  with bars showing the median~(P50) and whiskers spanning P10--P90~(log scale).}
  \Description{A single-column bar chart shows Conversation, Synthetic,
  ToolAgent, to-C, to-B API, Thinking, Coder, and Claude Code.
  Blue, orange, and green distinguish the Mooncake, Qwen Bailian, and Claude Code
  sources. Bar heights and numeric labels show median FP8 history KV footprint
  calculated from trace block counts in MiB per GPU rank under TP8, with P10--P90 error bars on a logarithmic
  vertical axis. Claude Code has a median footprint of 528 MiB per rank.}
  \label{fig:kv_footprint}
  \label{fig:motivation}
\end{figure}

The agentic coding trace~(Claude Code) exhibits the opposite reuse pattern
but reaches the same storage outcome.
Because each turn extends and replays the session prefix and sub-agent fan-out
reuses prefixes almost immediately~(mean reuse distance of 0.85 requests, P99 of 14),
only 16.6\% of blocks are one-shot and the infinite-capacity hit rate reaches 98.31\%.
However, each request spans an average of 3{,}421 blocks~(218{,}922 tokens),
compared with 23 blocks in Mooncake Conversation.
On DeepSeek-V4-Flash under FP8 and TP8~(Figure~\ref{fig:kv_footprint}),
the median Claude Code request carries a KV footprint of 528\,MiB per GPU rank~(4.1\,GiB across all eight ranks),
$12.1\times$ the largest median among the Mooncake and Qwen traces~(43.8\,MiB per rank),
while the P90 reaches 2{,}043\,MiB per rank~(16.0\,GiB total).
Across concurrent multi-turn sessions, the active working set rapidly exceeds host DRAM capacity.
Thus, whether reuse is sparse with long reuse distances or dense across multi-gigabyte contexts,
NVMe SSD must serve the bulk of cache hits,
so funneling all writes through DRAM adds migration without keeping hits in memory.

\begin{figure}[t]
  \centering
  \includegraphics[width=0.9\columnwidth]{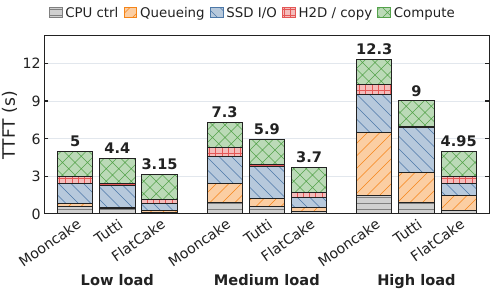}
  \caption{Measured TTFT breakdown of Mooncake, Tutti, and \sys under low, medium, and high load.
  Each bar separates TTFT into CPU control, queueing, SSD I/O, H2D/copy, and compute time,
  with the total TTFT~(in seconds) shown above the bar.
  Results are measured using DeepSeek-V4-Flash~(TP4) on 4$\times$B200 GPUs,
  replaying requests from the Claude Code trace, whose requests average 219K input tokens,
  at 0.3, 0.41, and 0.6\,QPS.}
  \Description{Nine stacked bars compare Mooncake, Tutti, and FlatCake across
  low, medium, and high load. Five segments show CPU control, queueing,
  SSD I/O, H2D/copy, and compute, with TTFT measured in seconds.
  Under high load, total TTFT is 12.3 seconds for Mooncake, 9.0 seconds for
  Tutti, and 4.95 seconds for FlatCake.}
  \label{fig:pipeline_stage_breakdown}
\end{figure}

\refstepcounter{obs}\label{obs:twohop}%
\noindent\textbf{Observation~\theobs: DRAM staging and eviction-ordered layout bottleneck SSD restoration.}
With SSD on the critical read path, end-to-end prefill latency is governed by the restoration pipeline.
Across low to high load, prefill compute stays constant at 2.0\,s
while non-compute restoration overhead~(CPU control, queueing, SSD I/O, and H2D copies) grows to 10.3\,s,
or 84\% of Mooncake's 12.3\,s high-load TTFT~(Figure~\ref{fig:pipeline_stage_breakdown}).
Although storage I/O accounts for 51--98\% of TTFT~\cite{impress2025},
raw NVMe bandwidth is not the primary limit,
since KV offloading achieves only 2.0\,GiB/s even when the SSD is emulated in DRAM at 16.9\,GiB/s~\cite{iochar2025}.
Instead, existing SSD restore paths bottleneck on host staging contention,
eviction-scrambled disk layout, or serialized SSD-only transfers.

First, staging SSD restores through host DRAM creates a concurrency bottleneck.
In a host-staged hierarchy~(Figure~\ref{fig:arch_comparison}(a)), restoring an SSD-resident prefix copies KV blocks from SSD
into host staging buffers before transferring them layer by layer to GPU HBM,
while background promotion separately populates the DRAM cache.
Even when chunked pipelining overlaps the $\text{SSD}\to\text{DRAM}$ and $\text{DRAM}\to\text{GPU}$ hops
to bound single-request transfer time near $\max(T_{\text{SSD}\to\text{DRAM}}, T_{\text{DRAM}\to\text{GPU}})$,
the inference engine must hold host staging buffers for each in-flight prefix until H2D transfers complete.
Staging memory demand therefore scales as $O(\text{context} \times \text{concurrency})$.
Under concurrent load, competing writes and restores exhaust the bounded staging pool,
which delays or aborts prefetches and forces recomputation of KV blocks
already stored on SSD, as \S\ref{sec:eval_e2e} shows.
As a result, Mooncake's queueing delay in Figure~\ref{fig:pipeline_stage_breakdown} rises from 0.2\,s at low load
to 5.0\,s at high load, or 41\% of its TTFT.

Second, bypassing host DRAM via batched GDS~\cite{spin2017,flashneuron2021,bam2023,geminifs2025,gofs2025,lmcache2025}
within a tiered hierarchy~(Figure~\ref{fig:arch_comparison}(a), as in Mooncake-GDS)
avoids host staging contention, but still reads an eviction-scrambled on-disk layout.
Tiered systems write blocks to SSD in DRAM eviction order across shared fixed-size FIFO buckets,
which scatters each session's prefix across non-contiguous SSD offsets.
Because each NVMe read command requires a contiguous LBA range on disk,
restoring a scattered prefix issues $O(\text{layers} \times \text{blocks})$ small reads,
consistent with a $3.3\times$ increase in SSD bandwidth yielding only a $1.33\times$ throughput gain~\cite{bidaw2026}.
Meanwhile, coarse FIFO bucket recycling evicts blocks from multiple sessions together
and invalidates active prefixes across sessions.

Finally, SSD-direct storage~(Tutti~\cite{tutti2025}, Figure~\ref{fig:arch_comparison}(b))
eliminates both cross-tier migration and host staging,
which cuts high-load H2D copy time from 0.8\,s to 0.1\,s~(Figure~\ref{fig:pipeline_stage_breakdown}).
Yet Tutti spends longer in SSD I/O than Mooncake at high load~(3.6\,s vs.\ 3.0\,s)
and still incurs a 9.0\,s TTFT.
Omitting host DRAM forces every cache hit, including compact shared prefixes, onto slower NVMe SSD,
while reading whole page objects that pack all transformer layers together~(\emph{page-major} layout)
serializes SSD transfers before layer computation can begin.

\noindent\textbf{Takeaway.}
Existing designs face these trade-offs because general-purpose tiering assumes mutable pages
whose future reuse and spatial grouping are unknown when written,
and therefore relies on DRAM migration and staging to manage locality.
In contrast, LLM KV cache blocks are \emph{write-once, immutable,
and reused together as a shared prompt or session prefix}.
Because blocks are never modified in place and both prefix sharing and write size are observable at creation time,
write-time placement can keep compact shared prefixes in host DRAM
and write large session histories sequentially to NVMe SSD in layer-major order,
where each layer's blocks occupy a contiguous on-disk range.
Managing DRAM and SSD as peer media thus eliminates migratory DRAM filtering
along with host staging and reassembly.

\section{The \sysn Design}
\label{sec:design}

\subsection{System Overview}
\begin{figure}[!t]
  \centering
  \includegraphics[width=\columnwidth]{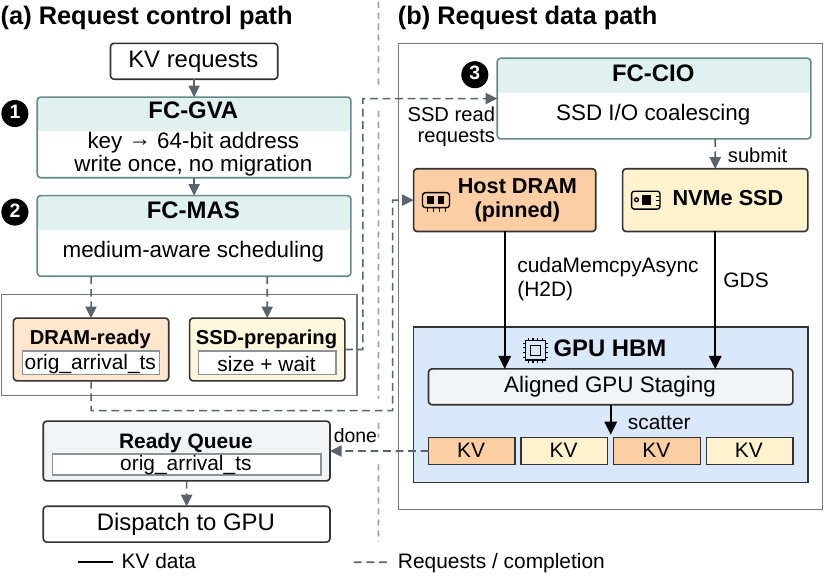}
  \caption{\sys architecture.
  (a)~Control path: \gva maps block keys to 64-bit GVAs with write-once placement~(\ding{182}),
  and \mas separates DRAM-ready and SSD-preparing paths into an arrival-ordered ready queue~(\ding{183}).
  (b)~Data path: DRAM hits use \texttt{cudaMemcpyAsync},
  while \cio streams coalesced KV data from SSD layer by layer
  via GDS~(\ding{184}) into an aligned GPU staging window.}
  \Description{Two-panel system overview showing the control path (FC-GVA mapping keys
  to 64-bit GVAs, FC-MAS separating DRAM-ready and SSD-preparing paths into a Ready Queue,
  and FC-CIO) and the data path (pinned host DRAM and NVMe SSD feeding an aligned GPU staging
  window and KV slots).}
  \label{fig:arch}
  \label{fig:gva_translation}
\end{figure}

\sys exposes host DRAM and NVMe SSD
as peer storage media behind a unified 64-bit Global Virtual Address~(GVA) space~(Figure~\ref{fig:arch}).
In a traditional tiered hierarchy, the intermediate DRAM layer implicitly handles three tasks:
selecting what spills to SSD via eviction, hiding SSD latency via pre-staging,
and reassembling scattered SSD blocks in host buffers.
Eliminating this intermediate layer removes cross-tier migration and staging contention,
but requires \sys to answer three design questions along the request lifecycle:

\rih{Q1: How to bind blocks to peer media at creation time with $O(1)$ routing?}
On the write path, \sys's \textbf{Global Virtual Address Manager}~(\textbf{\gva}, \ding{182}, \S\ref{sec:gva}) assigns each newly generated KV block
directly to DRAM or SSD once at creation time
and encodes its medium, node, and byte offset in an immutable 64-bit GVA.
On the read path, \gva resolves block keys to GVAs in $O(1)$ time without tier-by-tier lookup,
while unreferenced scopes are reclaimed in place without writeback.

\rih{Q2: How to schedule requests across heterogeneous media latencies without DRAM pre-staging?}
On the read control path, \sys's \textbf{Medium-Aware Request Scheduler}~(\textbf{\mas}, \ding{183}, \S\ref{sec:scheduler}) inspects the media bits in each GVA
to separate fast DRAM-ready requests from long-latency SSD-preparing requests.
DRAM-only hits enter an arrival-ordered ready queue directly.
Requests referencing SSD blocks prefetch asynchronously under a size-aware priority
and then join the same ready queue ordered by their original arrival timestamps,
so neither medium blocks or starves the other.

\rih{Q3: How to coalesce and stream SSD reads directly into a bounded GPU window?}
On the read data path, \sys's \textbf{Coalescing Direct I/O Engine}~(\textbf{\cio}, \ding{184}, \S\ref{sec:gds}) co-designs write-time layer-major scope layout
with read-time sort-merge coalescing to turn block-level prefix requests into sequential GPUDirect Storage~(GDS) transfers.
It streams each layer extent directly into a bounded GPU HBM window overlapped with layer-wise computation.

\subsection{\gva: Global Virtual Address Manager}
\label{sec:gva}
In a tiered hierarchy, the storage layer does not decide a block's long-term home at write time
because every block enters host DRAM first and LRU eviction later pushes cold blocks to SSD.
However, as Observations~\ref{obs:hitrate} and~\ref{obs:twohop} showed,
this reactive path pollutes DRAM with one-shot blocks and scatters session prefixes on disk.
Eliminating cross-tier migration therefore raises \textbf{Q1}:
\emph{Without reactive LRU eviction, how can \sys permanently bind KV blocks to DRAM or SSD at creation time
and route accesses in $O(1)$ time?}

\gva answers this question by exploiting two properties of LLM KV serving.
First, a request's sharing and size signals are observable upon arrival, before its KV blocks are written,
including whether its prefix is already shared, how many uncached tokens the turn adds,
and how large the session context has grown.
Second, flat peer storage bounds the cost of misplacement.
Because both media are directly accessible and SSD scopes are laid out contiguously on disk,
placing a reusable block on SSD causes neither a cache miss nor cross-tier migration,
and instead serves later hits sequentially via GDS at SSD bandwidth rather than DRAM bandwidth.

\rih{Write-Once Scope-Direct Placement}\label{sec:placement}
Rather than funneling all writes through DRAM, \gva assigns newly generated KV blocks directly to DRAM or SSD
at creation time using observable engine signals~(Table~\ref{tab:placement}).
Placement splits writes into two media classes matching the bimodal reuse regimes in Observation~\ref{obs:hitrate}.
\textbf{DRAM-bound writes (\texttt{MediaType=0})} cover globally shared system prompts~(radix-tree prefix reference count $>1$)
and short writes adding fewer than $\tau$ uncached tokens~\cite{bidaw2026},
including compact initial turns that serve as the shared base prefix reread by subsequent turns of a session.
Because these prefixes have small memory footprints and repeated reuse across requests or turns,
keeping them in host DRAM preserves sub-millisecond access across long inter-request intervals
without letting large one-shot writes flush them to disk.
\textbf{SSD-bound writes (\texttt{MediaType=1})} cover long writes adding at least $\tau$ uncached tokens,
long-context or agentic sessions once accumulated session context reaches $L_{\text{agent}}$,
and capacity spills when host DRAM is full.
We set $\tau = 8$K and $L_{\text{agent}} = 64$K tokens; when both rules apply, $L_{\text{agent}}$ takes precedence.
Blocks placed in DRAM are never migrated or demoted to SSD.
When host DRAM fills, new writes spill directly to SSD without generating DRAM-to-SSD eviction traffic,
while unreferenced scopes in either medium can be reclaimed in place without writeback.

Across turns of a multi-turn or agentic session, \gva evaluates placement per turn while keeping SSD blocks grouped by session.
If a session begins with a short DRAM-placed prompt and later appends a long turn~($\geq \tau$ uncached tokens)
or reaches $L_{\text{agent}}$, \gva directs the new turn's blocks to the session's SSD GPU File,
forming a hybrid DRAM--SSD prefix handled by \mas.
All SSD-bound turns of that session append contiguously to the same GPU File.
Across nodes, a coordinator tracks aggregate DRAM and SSD capacity and assigns GVAs on writes.
Because reads dominate KV traffic, capacity coordination stays off the critical read path.

\begin{table}[t]
  \centering
  \caption{\gva write-time placement rules. SSD rules take precedence once session context reaches $L_{\text{agent}}$ or host DRAM is full.}
  \label{tab:placement}
  \footnotesize
  \begin{tabular*}{\columnwidth}{@{\extracolsep{\fill}}lll@{}}
    \toprule
    Placement Case & Write-Time Signal & Target (\texttt{MediaType}) \\
    \midrule
    Shared prompt      & Prefix refcount $>1$                    & DRAM (\texttt{0}) \\
    Short write        & Uncached tokens $<\tau$                 & DRAM (\texttt{0}) \\
    Agentic / Long     & Session context $\geq L_{\text{agent}}$ & SSD (\texttt{1}) \\
    Long write / Spill & Uncached tokens $\geq\tau$ or DRAM full & SSD (\texttt{1}) \\
    \bottomrule
  \end{tabular*}
\end{table}

\rih{Unified 64-Bit Addressing and Routing}
Because a block never migrates after creation, its physical location remains invariant throughout its lifetime
and can be encoded directly in its address handle.
\gva translates each block key to a self-routing 64-bit address without tier-by-tier lookup~(Figure~\ref{fig:arch}(a)).
At \emph{Level~1}, a DRAM-resident hash table maps each block key~(session ID, layer, and token range)
to an immutable 64-bit GVA in a single $O(1)$ lookup.
To prevent read-after-write hazards without locking readers, \gva allocates a block's GVA before asynchronous write submission
but publishes the key-to-GVA entry into the Level~1 index only after the DMA or GDS write callback completes.
At \emph{Level~2}, the 64-bit GVA encodes three fixed bit fields:
\textbf{MediaType}~\texttt{[63:60]}~(4 bits, with \texttt{0} for DRAM, \texttt{1} for SSD, and reserved values for future media),
\textbf{NodeID}~\texttt{[59:48]}~(12 bits, up to 4{,}096 nodes),
and \textbf{Offset}~\texttt{[47:0]}~(48-bit byte offset, up to 256\,TiB per node).

Extracting \texttt{MediaType} and \texttt{NodeID} via register-level bit masks routes each access directly
to its transport backend without per-tier page tables or lock-protected LRU chains:
local DRAM via \texttt{cudaMemcpyAsync}, remote DRAM via RDMA,
local SSD via GDS \texttt{cuFile}, and remote SSD via an owner-side SSD read followed by an RDMA transfer.
Placing \texttt{MediaType} in the most significant bits also partitions DRAM and SSD addresses
into disjoint numeric ranges~($\mathrm{GVA} < 2^{60}$ for DRAM vs.\ $\mathrm{GVA} \geq 2^{60}$ for SSD),
so \mas can classify a request's medium with a single integer comparison.

\rih{Zero-Writeback Scope Reclamation}\label{sec:reclamation}
Write-once scope placement also simplifies storage reclamation.
Because KV blocks are never modified in place, freeing unreferenced scopes on either medium requires zero dirty-block writeback.
On SSD, \gva tracks active readers per session scope with an atomic reference count.
When SSD utilization crosses a high watermark, \sys reclaims the oldest unreferenced session extents
by returning their pre-allocated GPU Files to the free pool in $O(1)$ metadata operations.
In contrast, tiered systems such as Mooncake allocate SSD pages incrementally across fixed-size buckets
and recycle buckets in FIFO order, so evicting a single bucket slices across $k$ distinct sessions.
If affected session $i$ has re-access probability $p_i$,
a bucket eviction invalidates at least one active session prefix with probability
$1 - \prod_{i=1}^{k}(1 - p_i)$, which approaches certainty as $k$ grows and forces recomputation whenever an invalidated prefix is requested again.
Reclaiming whole session extents eliminates cross-session prefix invalidation
and preserves the contiguous on-disk layout required by \cio.

\subsection{\mas: Medium-Aware Request Scheduler}
\label{sec:scheduler}
Tiered systems hide SSD latency from the inference scheduler by pre-staging every SSD hit into host DRAM
before admission, which causes the $O(\text{context} \times \text{concurrency})$ host memory contention
identified in Observation~\ref{obs:twohop}.
Removing host DRAM pre-staging eliminates this memory bottleneck,
but exposes a $2.4\text{--}2.5\times$ bandwidth gap between local DRAM and NVMe SSD
directly to the request scheduler~(Table~\ref{tab:hardware}).
This raises \textbf{Q2}:
\emph{How can \sys prevent slow SSD restores from stalling fast DRAM hits without starving long-context SSD sessions?}

Scheduling directly across peer media creates a two-sided dilemma.
In a shared FCFS queue,
a multi-gigabyte SSD restore blocks subsequent DRAM-ready requests that need only a sub-millisecond \texttt{cudaMemcpyAsync}.
Conversely, strict DRAM-first priority scheduling can starve large SSD-backed agentic sessions under continuous DRAM arrivals.
\mas resolves both hazards through dual-queue separation, size-aware SSD prefetching,
and arrival-ordered re-admission into a unified ready queue.

\rih{Dual-Queue Separation}
When a request first arrives, \mas records its original request-arrival timestamp~($\text{orig\_arrival\_ts}$),
resolves its prefix block GVAs, and classifies the request by its slowest required medium~(Figure~\ref{fig:arch}(a)).
Requests whose cached blocks reside entirely in DRAM~($\max_j \mathrm{GVA}_j < 2^{60}$)
take the \textbf{DRAM-ready} path directly into the unified \textbf{Ready Queue} ordered by $\text{orig\_arrival\_ts}$,
from which requests are dispatched to the GPU.
Requests that reference any SSD-resident blocks~($\max_j \mathrm{GVA}_j \geq 2^{60}$)
enter the \textbf{SSD-Preparing Queue} instead,
including hybrid requests that combine a DRAM-placed prompt prefix with SSD-resident session turns.
For each preparing request, \mas pins any DRAM-resident prefix blocks in place
while \cio asynchronously prefetches the SSD-resident blocks into GPU HBM,
so SSD I/O overlaps with GPU execution on requests dispatched from the Ready Queue.

\rih{Size-Aware Prefetching and Arrival-Ordered Re-Admission}
\mas enforces fairness both within the SSD-Preparing Queue and across media at the Ready Queue.
First, to prevent large agentic restores from delaying smaller SSD requests on the NVMe array
without starving long-context sessions, the SSD-Preparing Queue orders pending prefetches
using a size-aware highest-response-ratio-next~(HRRN)~\cite{hrrn1971} priority
that substitutes SSD KV transfer volume for service time:
\begin{equation}
  \text{Priority}_i = 1 + \frac{\text{wait\_time}_i}{\text{KV\_size}_i}
  \label{eq:priority}
\end{equation}
Smaller SSD restores receive higher initial priority to reduce mean queuing delay,
whereas linear aging via $\text{wait\_time}_i$ guarantees eventual prefetch dispatch for multi-gigabyte restores.

Second, once \cio completes a request's SSD prefetch, its KV blocks are resident in GPU HBM.
Rather than appending the request to the tail of the Ready Queue,
\mas inserts it ordered by its \emph{original} arrival timestamp~($\text{orig\_arrival\_ts}$, Figure~\ref{fig:arch}(a)).
Because the time spent in SSD prefetching already counts toward its timestamp seniority,
the completed SSD request queues ahead of DRAM-ready requests that arrived during its prefetch,
so DRAM bursts cannot starve SSD-backed sessions.

\subsection{\cio: Coalescing Direct I/O Engine}
\label{sec:gds}
\begin{figure}[!t]
  \centering
  \includegraphics[width=0.9\columnwidth]{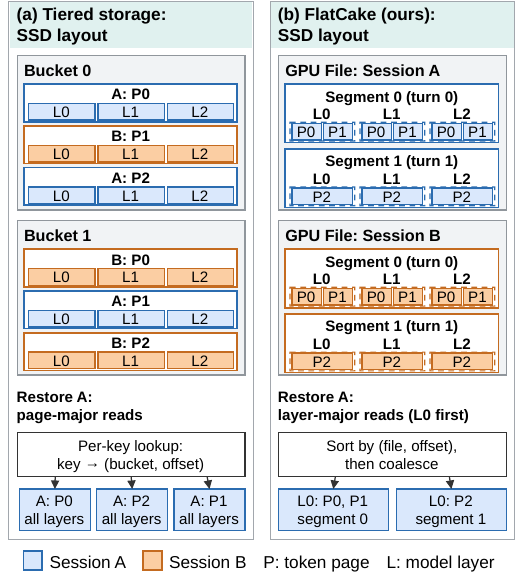}
  \caption{SSD layout and restore reads for the same KV data of two sessions.
  (a)~Tiered storage packs all-layer page objects from different sessions into shared buckets,
  so restoring session~A looks up and reads each page object separately (page-major).
  (b)~\sys assigns each session its own GPU File where each turn appends a layer-major segment.
  Restoring session~A sorts and coalesces block descriptors into one contiguous range
  per layer per segment, starting with layer~0 (layer-major).}
  \Description{Two side-by-side panels over the same KV data of sessions A (blue) and B (orange),
  each with three token pages P0 to P2 and three layers L0 to L2.
  (a) Tiered storage: Bucket 0 and Bucket 1 each hold page objects from both sessions; every page
  object contains L0, L1 and L2. Restoring session A looks up each key to a bucket and offset and
  reads three all-layer page objects.
  (b) FlatCake: one GPU File per session; Segment 0 (turn 0) holds P0 and P1, Segment 1 (turn 1)
  holds P2, and within each segment the pages of each layer are grouped in a dashed contiguous range.
  Restoring session A sorts descriptors by file and offset and coalesces them into two layer-0 reads,
  one per segment.}
  \label{fig:kvcache_layout}
\end{figure}

\begin{figure}[!t]
  \centering
  \includegraphics[width=0.9\columnwidth]{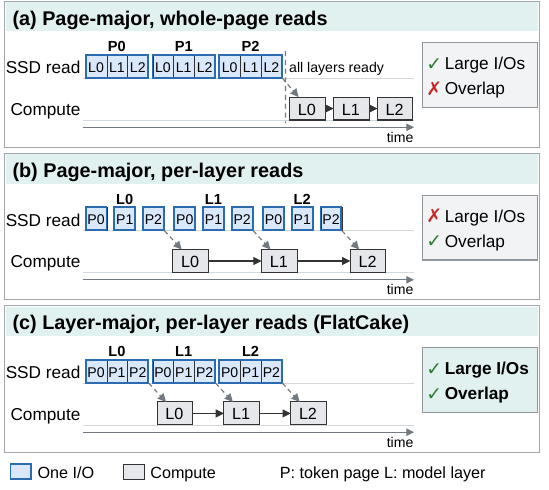}
  \caption{Read granularity versus read/compute overlap for the same KV bytes
  (3 pages $\times$ 3 layers).
  (a)~Page-major whole-page reads issue large I/Os, but every page object carries all layers,
  so computation waits for the last object.
  (b)~Page-major per-layer range reads overlap computation,
  but issue one small, non-coalescible I/O per page per layer.
  (c)~\sys's layer-major layout issues one contiguous I/O per layer
  and reads layer~$l{+}1$ while computing layer~$l$.
  }
  \Description{Three stacked timelines, each with an SSD read lane and a compute lane.
  (a) Three large reads, one per page object containing layers L0 to L2, finish before
  layer computation starts; marked large I/Os yes, overlap no.
  (b) Nine small reads, one per page per layer, grouped by layer; each layer's computation
  starts after its three reads; marked large I/Os no, overlap yes.
  (c) Three large reads, one per layer containing pages P0 to P2; computation of each layer
  overlaps the read of the next layer and finishes first; marked large I/Os yes, overlap yes.}
  \label{fig:layer_overlap}
\end{figure}

Removing host DRAM staging eliminates host memory contention,
but also removes the host buffer that previously reassembled scattered SSD reads.
This raises \textbf{Q3}:
\emph{Without host DRAM buffers, how can \sys turn block-level prefix requests into sequential NVMe transfers
and stream them into a bounded GPU HBM budget?}
Because NVMe DMA requires a contiguous source LBA range to coalesce a transfer~(Observation~\ref{obs:twohop}),
\cio co-designs write-time disk layout, read-time descriptor merging,
and layer-pipelined GPU streaming~(Figures~\ref{fig:kvcache_layout} and~\ref{fig:layer_overlap}).

\rih{Write-Time Layer-Major Scope Layout}
During prefill, transformer layer~$l$ consumes all cached prefix tokens for layer~$l$
before execution advances to layer~$l{+}1$.
Page-major SSD layouts mismatch this consumption order under direct I/O.
When tiered storage packs all-layer page objects from concurrent sessions into shared eviction buckets~(Figure~\ref{fig:kvcache_layout}(a)),
whole-page reads preserve large I/O sizes but stall layer~0 computation until the final page object arrives~(Figure~\ref{fig:layer_overlap}(a)).
Per-layer range reads into those page objects restore compute overlap,
but split each restore into $O(\text{layers} \times \text{blocks})$ small, non-coalescible reads across scattered offsets~(Figure~\ref{fig:layer_overlap}(b)).
\cio resolves both halves of this dilemma by mapping each SSD-bound session scope to a dedicated \emph{GPU File}
backed by a pre-allocated, single-extent file in an NVMe file pool~(Figure~\ref{fig:kvcache_layout}(b)).
Within a session's GPU File, each SSD-bound turn appends a \emph{layer-major segment}
in which all token blocks of layer~$l$ occupy one physically contiguous byte range before layer~$l{+}1$.
Restoring a prefix therefore issues $O(\text{layers})$ contiguous reads per segment
while overlapping layer~$l{+}1$ GDS reads with layer~$l$ computation~(Figure~\ref{fig:layer_overlap}(c)).

\rih{Read-Time Sort-Merge Coalescing}
Even when blocks are contiguous on disk, the inference engine's radix tree queries prefixes at page granularity
and hands the storage backend hundreds of thousands of block descriptors that may span multiple turns or scopes
and arrive out of offset order.
To recover contiguous LBA ranges before issuing GDS calls,
\cio performs a sort-merge pass at the submission layer~(Figure~\ref{fig:kvcache_layout}(b)).
It first sorts block descriptors by \texttt{(gpu\_file\_id, offset)} to place physically adjacent blocks next to one another,
and then scans linearly to coalesce adjacent ranges into single large read descriptors.
When partial prefix hits or sub-block alignments leave small unrequested gaps between adjacent ranges,
\cio bridges gaps smaller than a configurable byte threshold~($\Delta_{\text{gap}} = 64$\,KiB),
which trades a bounded amount of extra sequential I/O to avoid splitting NVMe commands.
Merged descriptors are striped round-robin across worker threads,
and each thread submits batches via a private \texttt{cuFileHandle}, CUDA stream, and \texttt{cuFileBatchIOSubmit}.

\rih{Layer-Pipelined Direct GPU Streaming}
Because SGLang's paged KV cache allocates 256-token HBM pages non-contiguously,
issuing direct GDS reads into individual KV page slots would split each layer extent back into per-page DMA transfers.
Conversely, staging an entire multi-gigabyte prefix in host DRAM causes $O(\text{context} \times \text{concurrency})$ memory contention.
\cio avoids both pitfalls by streaming coalesced layer extents
via GDS into a double-buffered GPU HBM staging window~(sized to two transformer layers per active stream)
before scattering blocks into paged HBM KV slots on the GPU~(Figure~\ref{fig:arch}(b)).
While the compute stream executes attention and FFN kernels for layer~$l$,
the I/O stream asynchronously executes \texttt{cuFileBatchIOSubmit} and a GPU scatter kernel for layer~$l{+}1$~(Figure~\ref{fig:layer_overlap}(c)).
This pipeline bounds staging memory to a constant $O(1)$ HBM window per GPU.
When concurrent restores saturate the window under heavy load, incoming prefetches wait
on CUDA stream events rather than revoking in-flight prefetches and recomputing cached tokens.

\section{Implementation}
\label{sec:impl}
We implement \sys in approximately 14K lines of C++20/CUDA and 7K lines of Python,
integrating it into SGLang~\cite{sglang2024} behind the standard \texttt{HiCacheStorage} interface~(\texttt{exists}, \texttt{get}, \texttt{put})
without modifying model runner or attention kernels.

\rih{Pinned DRAM pool and hugepage allocation}
At engine initialization, the local DRAM backend pre-allocates the host KV pool using \texttt{mmap}
with \texttt{MAP\_POPULATE} and 2\,MB hugepages~(\texttt{MAP\_HUGETLB}),
and registers the arena once with \texttt{cudaHostRegister}.
Pre-faulting and pinning hugepages at startup eliminates page-fault traps, TLB thrashing,
and per-transfer pinning overhead on the \texttt{cudaMemcpyAsync} path.

\rih{Pre-allocated NVMe file pool and GDS alignment}
To prevent filesystem metadata allocation and extent fragmentation on the write path,
\cio pre-allocates a pool of fixed-size single-extent files on the XFS/RAID-0 array at startup using \texttt{fallocate}
and opens each file with \texttt{O\_DIRECT} and \texttt{cuFileHandleRegister}.
All logical block offsets within a GPU File and all GPU staging buffer slices are aligned to 4\,KiB boundaries,
satisfying \texttt{libcufile} DMA alignment requirements and preventing driver-internal bounce buffering.

\rih{Double-buffered GPU staging and page scatter}
Because calling \texttt{cuFileBufRegister} dynamically on per-request GPU buffers incurs millisecond-scale
BAR1 and IOMMU mapping overhead, \cio allocates and registers a fixed per-GPU HBM staging window once at initialization.
Restores pipeline GDS reads across two ping-pong layer buffers within this registered window;
as soon as a coalesced layer read completes on the I/O CUDA stream,
a lightweight CUDA scatter kernel copies each 256-token page from the contiguous staging buffer
into its non-contiguous destination slot in SGLang's paged KV pool,
synchronizing with the model compute stream via \texttt{cudaEventRecord} and \texttt{cudaStreamWaitEvent}.

\rih{Lock-free Level-1 GVA index}
The Level~1 block-key-to-GVA index is implemented as a concurrent open-addressed hash table
using 64-bit atomic compare-and-swap~(\texttt{CAS}) operations.
Because GVAs are immutable once published and session reclamation frees entire scopes asynchronously,
scheduler threads querying prefix hits execute wait-free reads without contending with background write-completion workers.

\section{Evaluation}
\label{sec:eval}
We evaluate \sys across a 284B-parameter hybrid MoE model on NVIDIA B200 GPUs
and a 70B-parameter GQA model on NVIDIA A100 GPUs to answer three questions.
\textbf{(E1)}~Does replacing tiered DRAM staging with flat peer storage prevent cache hit-rate collapse under load,
and how does this affect TTFT, TPOT, and total token throughput across different reuse regimes~(\S\ref{sec:eval_e2e})?
\textbf{(E2)}~How does \gva distribute delivered hits across GPU HBM, host DRAM, and NVMe SSD while reducing SSD write traffic,
how does \cio improve SSD read bandwidth and read amplification,
and how does \mas coordinate mixed-medium reuse~(\S\ref{sec:eval_gains})?
\textbf{(E3)}~Do \sys's layout and streaming gains hold on standard grouped-query attention across context lengths up to 128K tokens,
and can direct GDS transport alone match flat layer-major storage~(\S\ref{sec:eval_sensitivity})?

\subsection{Experimental Setup}
\label{sec:eval_setup}
\rih{Testbed}
Our primary evaluation runs on 4 GPUs of an 8-GPU NVIDIA B200 server with 3\,TB of host DRAM
and an 8-drive Samsung PM9A3 NVMe SSD RAID~0 array~(Table~\ref{tab:hardware}).
Systems with external KV caching share a capacity budget of up to 512\,GiB host DRAM and 2\,TiB NVMe SSD,
each using only the media its architecture supports.
We build on SGLang v0.5.19, Mooncake v0.3.13, CUDA 13.1, and NVIDIA GPUDirect Storage~(GDS) 1.16.1~(\texttt{libcufile}).
The cross-model scaling experiments in \S\ref{sec:eval_sensitivity} run on 4 GPUs of an 8-GPU NVIDIA A100 server.

\begin{table}[t]
  \centering
  \caption{Hardware and measured GPU-path bandwidth.}
  \label{tab:hardware}
  \label{tab:access_paths}
  \footnotesize
  \setlength{\tabcolsep}{4pt}
  \begin{tabular*}{\columnwidth}{@{\extracolsep{\fill}}llrr@{}}
    \toprule
    \textbf{Component} & \multicolumn{3}{l@{}}{\textbf{Configuration}} \\
    \midrule
    GPU           & \multicolumn{3}{l@{}}{8$\times$NVIDIA B200, 180\,GB HBM3e each} \\
    Interconnect  & \multicolumn{3}{l@{}}{NVLink via NVSwitch (18 links/GPU), PCIe Gen5} \\
    CPU           & \multicolumn{3}{l@{}}{2$\times$Xeon Platinum 8570 (112 cores, 4 NUMA nodes)} \\
    Host DRAM     & \multicolumn{3}{l@{}}{3\,TB} \\
    NVMe SSD      & \multicolumn{3}{l@{}}{8$\times$3.84\,TB Samsung PM9A3, RAID~0} \\
    \midrule
    \textit{GPU path} & \textit{Mechanism} & \textit{Read (GiB/s)} & \textit{Write (GiB/s)} \\
    \cmidrule{1-4}
    Local DRAM            & \texttt{cudaMemcpyAsync}         & 46--48  & 32--34 \\
    Local SSD             & \texttt{cuFile} (GDS)            & 19.3    & 21.3 \\
    Remote DRAM           & GPUDirect RDMA                   & 31--33  & 28--30 \\
    Remote SSD            & owner read + RDMA                & 6.3     & 7.8 \\
    \bottomrule
  \end{tabular*}
\end{table}

\rih{Model and workload}
We serve DeepSeek-V4-Flash-0731~\cite{deepseekv4} with TP4, a 284B-parameter MoE model
with 43 hybrid multi-head latent attention~(MLA) layers and a 1M-token context window.
The model uses an \texttt{fp8\_e4m3} KV cache with 256-token pages and a per-rank footprint of 7.5\,KiB/token.
We replay three production KV cache traces from Table~\ref{tab:hitrate} that span both reuse regimes identified in Observation~\ref{obs:hitrate},
covering multi-gigabyte agentic coding~(\texttt{claude-\allowbreak code-\allowbreak weka-\allowbreak 0621}~\cite{ccTracesWeka2026})
as well as code generation~(Qwen Bailian Coder~\cite{kvcache_wild2026})
and multi-turn conversation~(Mooncake Conversation~\cite{mooncake2024}) with high once-access ratios and long reuse distances.
Requests are replayed in recorded arrival order without cache warm-up at 0.2--0.6\,QPS~(Claude Code),
5--12\,QPS~(Qwen Bailian), and 1.5--4.5\,QPS~(Mooncake),
following the SGLang HiCache benchmark methodology~\cite{sglang_hicache2025}.
To evaluate generality across model architectures and context lengths,
we also serve Llama-3.1-70B~(dense GQA, TP4) with input lengths from 8K to 128K tokens~(\S\ref{sec:eval_sensitivity}).

\begin{figure}[!t]
    \centering
    \includegraphics[width=\columnwidth]{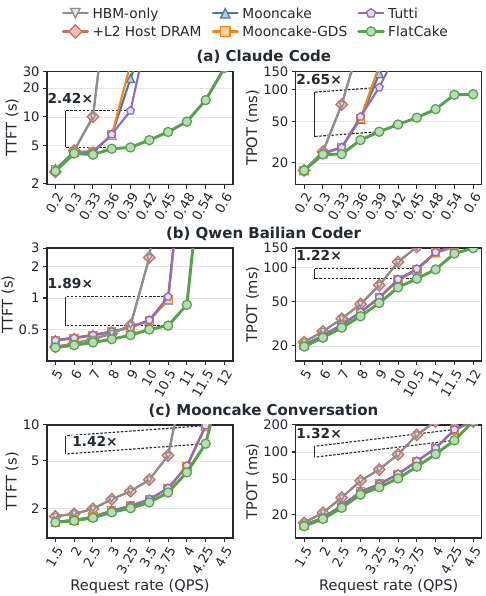}
    \caption{Mean TTFT~(left) and TPOT~(right) across request rates on (a)~Claude Code,
    (b)~Qwen Bailian Coder, and (c)~Mooncake Conversation (log scale, clipped).
    Brackets show \sys's speedup over Tutti at 0.39, 10.5, and 4.25\,QPS.}
    \Description{A three-by-two grid of line charts compares HBM-only, +L2 Host DRAM,
    Mooncake, Mooncake-GDS, Tutti, and FlatCake. Rows (a), (b), and (c) correspond to
    Claude Code, Qwen Bailian Coder, and Mooncake Conversation; the left column plots
    mean TTFT and the right column plots mean TPOT against evenly spaced request rates on
    logarithmic y-axes. TTFT display limits are 30, 3, and 10 seconds, respectively; TPOT
    display limits are 150, 150, and 200 milliseconds. Both columns use dashed guides to compare
    Tutti and FlatCake at 0.39, 10.5, and 4.25 QPS for the three traces, respectively.
    Each ratio is Tutti's mean TTFT or TPOT divided by FlatCake's corresponding latency
    at the same request rate.}
    \label{fig:serving_latency}
\end{figure}

\rih{Baselines and metrics}
We compare \sys against five baselines on the same inference engine, hardware, and storage capacity budgets.
\textbf{HBM-only} uses SGLang's native GPU HBM KV cache without external storage,
and \textbf{+L2 Host DRAM}~\cite{sglang_hicache2025} adds SGLang HiCache's two-tier HBM--DRAM LRU cache without SSD offloading.
Among the SSD-backed baselines, \textbf{+L3 Mooncake}~\cite{mooncake2024} implements a three-tier HBM--DRAM--SSD hierarchy
that stages SSD-resident blocks through host DRAM before H2D transfer and promotes them into DRAM,
\textbf{+L3 Mooncake-GDS} replaces host-staged SSD reads in Mooncake with direct GDS transfers while retaining its migratory DRAM tier,
and \textbf{Tutti}~\cite{tutti2025} transfers KV blocks directly between GPU HBM and NVMe SSD via GPU-executed I/O queues while bypassing host DRAM.

\subsection{Overall Performance}
\label{sec:eval_e2e}
\begin{figure}[!t]
    \centering
    \includegraphics[width=\columnwidth]{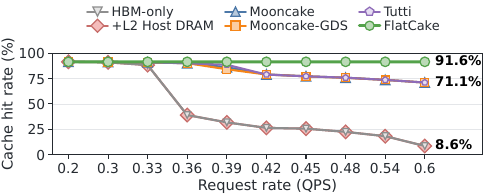}
    \caption{Delivered cache hit rate on Claude Code.}
    \Description{A line chart compares cache hit rates of HBM-only, +L2 Host DRAM, Mooncake,
    Mooncake-GDS, Tutti, and FlatCake on the Claude Code trace across ten evenly spaced
    request rates from 0.2 to 0.6 QPS, on a zero-to-100 percent vertical axis.
    FlatCake stays at 91.6 percent at every rate; Mooncake, Mooncake-GDS, and Tutti
    decline to about 71 percent, and HBM-only and +L2 Host DRAM fall to 8.6 percent.}
    \label{fig:cache_hit_rate}
\end{figure}

\begin{figure}[!t]
    \centering
    \includegraphics[width=0.9\columnwidth]{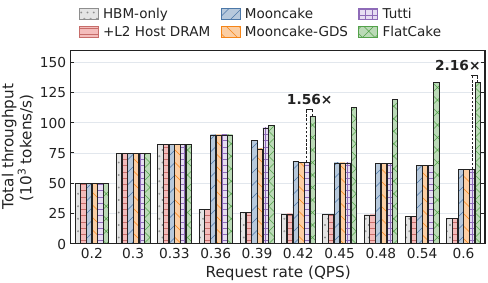}
    \caption{Total token throughput on Claude Code.
    Brackets show \sys's speedup over Tutti at 0.42\,QPS (past Tutti's knee) and 0.6\,QPS (maximum).}
    \Description{A grouped bar chart compares HBM-only, +L2 Host DRAM, Mooncake,
    Mooncake-GDS, Tutti, and FlatCake on the Claude Code trace. Ten QPS groups from
    0.2 to 0.6 each contain six bars; bar heights show total token throughput in
    thousands of tokens per second on a zero-based vertical axis.}
    \label{fig:throughput}
\end{figure}

To answer \textbf{E1}, Figures~\ref{fig:serving_latency}--\ref{fig:throughput} evaluate end-to-end serving latency,
delivered cache hit rate, and token throughput as request load scales across the three production traces.

\rih{Time to First Token}
\sys matches the fastest baseline at low load and reduces mean TTFT near the baselines' latency knees
by $1.40\times$--$2.42\times$ over the fastest baseline~($1.42\times$--$2.42\times$ over Tutti, Figure~\ref{fig:serving_latency}, left).
On the multi-gigabyte agentic workload~(Claude Code, Figure~\ref{fig:serving_latency}(a)),
all systems achieve similar TTFT at 0.2\,QPS~(2.70\,s for \sys vs.\ 2.76\,s for the fastest baseline)
when active sessions fit in GPU HBM.
As concurrency rises, multi-gigabyte session histories overflow HBM and the 512\,GiB DRAM tier,
causing HBM-only and +L2 Host DRAM to hit their latency knees first at 0.36\,QPS~(437.5\,s).
The SSD-backed baselines~(Mooncake, Mooncake-GDS, and Tutti) hit their knees next at 0.39\,QPS,
where the fastest baseline~(Tutti) reaches 11.60\,s while \sys maintains 4.79\,s~($2.42\times$ lower).
On Qwen Bailian Coder and Mooncake Conversation~(Figures~\ref{fig:serving_latency}(b)--(c)),
where 61.0--75.8\% of blocks are one-shot and reusable prefixes span hundreds of intervening requests~(Table~\ref{tab:hitrate}),
\sys still delays the latency knee by avoiding one-shot DRAM migration and fragmented SSD reads.
At 10.5\,QPS on Qwen Bailian Coder, \sys achieves a 0.544\,s mean TTFT,
which is $1.76\times$ lower than the fastest baseline~(Mooncake at 0.956\,s) and $1.89\times$ lower than Tutti.
At 4.25\,QPS on Mooncake Conversation, \sys achieves 6.97\,s,
outperforming Mooncake-GDS~(9.753\,s) by $1.40\times$ and Tutti by $1.42\times$.

\rih{Why the Baselines Degrade: Delivered Cache Hit Rate Across Regimes}
Comparing delivered cache hit rates across the two workload regimes explains why the baselines hit an early latency wall.
On Claude Code, \sys sustains a 91.6\% delivered hit rate from 0.2 to 0.6\,QPS,
whereas Mooncake, Mooncake-GDS, and Tutti drop to 71.1\%, and HBM-only and +L2 Host DRAM fall to 8.6\%~(Figure~\ref{fig:cache_hit_rate}).
In Mooncake, concurrent multi-gigabyte restores contend for bounded host DRAM staging buffers~(Observation~\ref{obs:twohop}),
which delays or aborts prefetches and forces the engine to recompute KV blocks already stored on SSD.
Although Tutti bypasses host DRAM, its page-major layout cannot overlap SSD reads with layer-wise computation~(Figure~\ref{fig:layer_overlap}(a)),
so concurrent restores still miss their prefill deadlines and fall back to recomputation.
Because every missed token must be recomputed on the GPU, this 20.5-point hit-rate loss inflates prefill execution time
and compounds queuing delay across waiting requests.
In contrast, on the lighter Qwen Bailian Coder and Mooncake Conversation traces,
delivered hit rates remain within 2--3 percentage points up to the latency knees,
where \sys reaches 60.0\% versus 57.3--57.7\% for the SSD-backed baselines at 10.5\,QPS on Qwen Bailian Coder
and 33.5\% versus 31.3--31.4\% at 4.25\,QPS on Mooncake Conversation, widening to at most 9.5 points beyond those rates.
Where hit rates barely diverge before saturation, \sys's TTFT reductions come directly from \gva's write-once placement
and \cio's sequential layer-major streaming rather than differences in cache retention.

\rih{Time per Output Token}
By preventing fallback recomputation and SSD restore stalls during prefill,
\sys also keeps elongated prefill batches from delaying interleaved decode iterations under continuous batching~(Figure~\ref{fig:serving_latency}, right).
Below the latency knee, \sys matches the fastest baseline at 17.0 versus 16.9\,ms on Claude Code at 0.2\,QPS.
Near the baselines' latency knees, \sys reduces mean TPOT over the fastest baseline by
$2.65\times$ on Claude Code at 0.39\,QPS~(39.8 vs.\ 105.4\,ms for Tutti),
$1.20\times$ on Qwen Bailian Coder at 10.5\,QPS~(79.9 vs.\ 95.6\,ms for Mooncake, and $1.22\times$ over Tutti),
and $1.30\times$ on Mooncake Conversation at 4.25\,QPS~(134.7 vs.\ 175.7\,ms for Mooncake-GDS, and $1.32\times$ over Tutti),
before all systems saturate near 219\,ms at 4.5\,QPS on Mooncake Conversation.

\rih{Total Token Throughput}
Sustaining a 91.6\% delivered hit rate on Claude Code allows \sys to scale total token throughput
up to $2.16\times$ beyond the SSD-backed baselines~(Figure~\ref{fig:throughput}).
All systems match the offered load up to 0.36\,QPS, where HBM-only and +L2 Host DRAM collapse to 28K\,tokens/s.
Past Tutti's knee at 0.39\,QPS, recomputation of missed prefixes caps Mooncake, Mooncake-GDS, and Tutti at 61--68K\,tokens/s.
Because \sys streams SSD prefixes into a bounded GPU window without staging aborts or recomputation,
it continues scaling to 133.1K\,tokens/s at 0.54\,QPS,
achieving $1.56\times$ Tutti's throughput at 0.42\,QPS and $2.16\times$ at 0.6\,QPS.

\subsection{Mechanism-Level Analysis}
\label{sec:eval_gains}
\begin{figure}[!t]
  \centering
  \includegraphics[width=0.85\columnwidth]{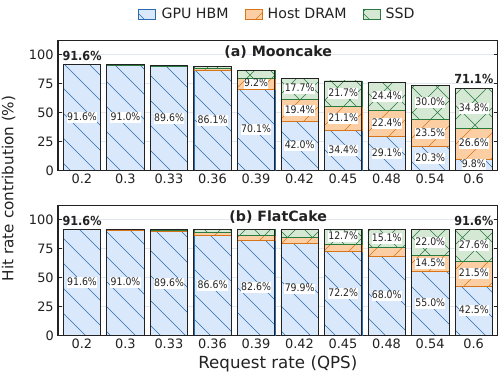}
  \caption{Delivered cache hit rate on Claude Code by serving medium for (a)~Mooncake and (b)~\sys.}
  \Description{Two stacked bar charts, Mooncake above FlatCake, show GPU HBM, Host DRAM, and SSD
  hit contributions at ten request rates from 0.2 to 0.6 QPS.
  Both systems start at 91.6 percent, entirely from HBM.
  At 0.6 QPS, Mooncake has 9.8 percent HBM, 26.6 percent DRAM, and 34.8 percent SSD contributions,
  for a total of 71.1 percent. FlatCake has 42.5 percent HBM, 21.5 percent DRAM,
  and 27.6 percent SSD contributions, maintaining a total of 91.6 percent.
  Small segments are plotted even where their numerical labels do not fit.}
  \label{fig:hit_composition}
\end{figure}

\begin{figure}[!t]
  \centering
  \includegraphics[width=\columnwidth]{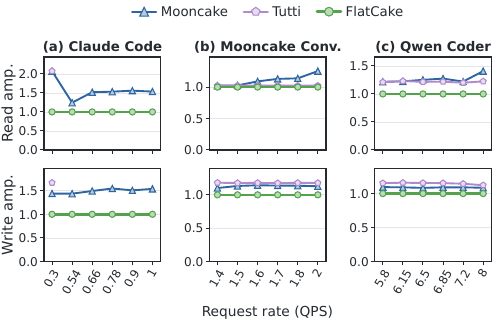}
  \vspace{-5mm}
  \caption{SSD read amplification~(top: KV tokens fetched from SSD per fetched token used) and
  write amplification~(bottom: SSD bytes written, normalized to \sys{}${}=1.0\times$)
  on (a)~Claude Code, (b)~Mooncake Conversation, and (c)~Qwen Bailian Coder; lower is better.}
  \Description{Six line charts in two rows and three columns compare Mooncake, Tutti, and FlatCake.
  Columns are Claude Code, Mooncake Conversation, and Qwen Bailian Coder. The top row shows read
  amplification, the ratio of KV tokens fetched from SSD to fetched tokens actually used; FlatCake is
  1.0 everywhere, Mooncake is 1.24 to 2.07 on Claude Code, 1.02 to 1.25 on Mooncake Conversation,
  and 1.22 to 1.41 on Qwen Bailian Coder, and Tutti is 1.02 and 1.20 to 1.23 on the last two traces.
  The bottom row shows SSD bytes written relative to FlatCake: Mooncake writes 1.44 to 1.55 times
  on Claude Code and 1.08 to 1.14 times on the other traces, and Tutti 1.12 to 1.18 times on the latter two traces.
  Tutti has one Claude Code point at 0.3 QPS, with read amplification of 2.07 times
  and normalized write amplification of 1.67 times.}
  \label{fig:io_amplification}
\end{figure}

\begin{figure}[!t]
  \centering
  \includegraphics[width=0.85\columnwidth]{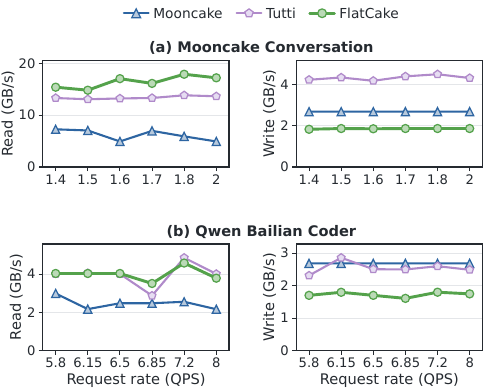}
  \caption{Peak SSD read~(left) and write~(right) bandwidth on
  (a)~Mooncake Conversation and (b)~Qwen Bailian Coder.}
  \Description{Four line charts in a two-by-two grid compare Mooncake, Tutti, and FlatCake
  on DeepSeek-V4-Flash with B200 TP4. The top row, panel (a), shows Mooncake Conversation
  at six request rates from 1.4 to 2 QPS. The bottom row, panel (b), shows Qwen Bailian Coder
  at six request rates from 5.8 to 8 QPS. Left panels report peak SSD read bandwidth
  and right panels report peak SSD write bandwidth in GB/s.
  Mooncake uses blue triangles, Tutti purple pentagons, and FlatCake green circles.
  All vertical axes start at zero.}
  \label{fig:mechanism}
\end{figure}

To answer \textbf{E2}, we analyze how \gva, \cio, and \mas contribute to \sys's performance
through per-medium hit attribution~(Figure~\ref{fig:hit_composition}),
SSD read/write amplification~(Figure~\ref{fig:io_amplification}),
peak SSD bandwidth~(Figure~\ref{fig:mechanism}),
and mixed-medium scheduling behavior.

\rih{\gva: Per-medium hit retention and recovery}
\label{sec:eval_placement}
\gva's flat peer placement and \cio's direct restoration preserve GPU HBM locality under load
while recovering all evicted HBM hits across host DRAM and NVMe SSD~(Figure~\ref{fig:hit_composition}).
At 0.2\,QPS on Claude Code, both \sys and Mooncake achieve a 91.6\% hit rate entirely from GPU HBM.
As arrival rate increases to 0.6\,QPS, Mooncake's HBM hit contribution drops to 9.8\%,
whereas \sys still serves 42.5\% of prompt tokens directly from HBM under the same HBM capacity.
Because Claude Code exhibits immediate turn-to-turn reuse~(mean reuse distance of 0.85 requests, Table~\ref{tab:hitrate}),
completing requests $2.42\times$ faster in \sys allows active sessions to issue their next turn while their prefix remains resident in HBM,
whereas stalled restores and recomputation in Mooncake accumulate in-flight requests that thrash HBM.
For the 49.1 percentage points of hits that do spill out of HBM at 0.6\,QPS,
\sys recovers the entire difference through host DRAM~(21.5\%) and NVMe SSD~(27.6\%) to maintain a 91.6\% delivered hit rate.
In contrast, Mooncake recovers only 61.4 of its 81.8 lost HBM points across DRAM~(26.6\%) and SSD~(34.8\%)
because host staging contention revokes or truncates in-flight restores under concurrency.

\rih{\gva: Write-once placement and SSD write reduction}
By binding each block to DRAM or SSD once at creation time, \gva eliminates DRAM-to-SSD eviction traffic
and reduces total SSD bytes written across all three traces~(Figure~\ref{fig:io_amplification}, bottom row).
Using \sys's measured SSD write volume at each trace and request rate as the $1.0\times$ reference~(with no SSD capacity reclamation triggered),
Mooncake writes $1.44\times$--$1.55\times$ as much SSD data on Claude Code,
$1.10\times$--$1.14\times$ on Mooncake Conversation, and $1.08\times$--$1.10\times$ on Qwen Bailian Coder
because one-shot and short-lived blocks enter DRAM first and later spill to SSD.
Tutti avoids cross-tier migration by bypassing host DRAM,
yet it still writes $1.18\times$ as much data to SSD on Mooncake Conversation
and $1.12\times$--$1.16\times$ on Qwen Bailian Coder
because it sends every KV write to NVMe SSD, whereas \gva absorbs shared prompts and short writes in host DRAM.

\rih{\cio: Layer-major SSD bandwidth and read amplification}
\label{sec:eval_io}
\cio's session-contiguous, layer-major layout and read-time sort-merge coalescing sustain higher sequential SSD read bandwidth
while eliminating wasted SSD reads from dropped prefetches~(Figures~\ref{fig:io_amplification} and~\ref{fig:mechanism}).
On Mooncake Conversation~(Figure~\ref{fig:mechanism}(a)), \sys reaches 14.84--17.92\,GB/s peak SSD read bandwidth,
achieving $2.10\times$--$3.49\times$ Mooncake and $1.13\times$--$1.29\times$ Tutti at matching request rates
because \cio's layer-major extents coalesce each restore into large sequential transfers,
whereas Mooncake stages scattered reads through host buffers and Tutti issues page-major reads.
On Qwen Bailian Coder~(Figure~\ref{fig:mechanism}(b)), whose shorter 5.7K-token prompts leave fewer blocks to coalesce,
\sys reaches 3.51--4.57\,GB/s, outperforming Mooncake by $1.35\times$--$1.87\times$ while matching Tutti~($0.94\times$--$1.22\times$).
Because \sys writes only SSD-placed scopes to disk, it also incurs the lowest peak SSD write bandwidth across both traces
at 1.83--1.87\,GB/s on Mooncake Conversation and 1.61--1.80\,GB/s on Qwen Bailian Coder~(Figure~\ref{fig:mechanism}, right).
In the top row of Figure~\ref{fig:io_amplification}, read amplification measures KV tokens fetched from SSD per fetched token actually used.
Because \cio streams coalesced extents directly into a bounded GPU window and waits on CUDA stream events rather than revoking in-flight transfers,
\sys never discards a fetched block~($1.00\times$).
In contrast, Mooncake fetches $1.24\times$--$2.07\times$ as many SSD tokens per used token on Claude Code,
$1.02\times$--$1.25\times$ on Mooncake Conversation, and $1.22\times$--$1.41\times$ on Qwen Bailian Coder
because host-staged prefetches are dropped when staging contention prevents them from being spliced onto the request prefix in time,
while Tutti fetches $1.02\times$ and $1.20\times$--$1.23\times$ on the latter two traces.

\rih{\mas: Scheduling implications of mixed-medium reuse}
\label{sec:eval_hol}
Figure~\ref{fig:hit_composition}(b) shows that flat peer storage serves a substantial share of hits from both media simultaneously under load,
with host DRAM and NVMe SSD contributing 21.5\% and 27.6\% of prompt tokens at 0.6\,QPS on Claude Code.
Combined with the $2.4\text{--}2.5\times$ local read-bandwidth gap between DRAM and SSD~(Table~\ref{tab:access_paths}),
this concurrent mix requires decoupling fast DRAM hits from long-latency SSD restores.
As detailed in \S\ref{sec:scheduler}, \mas dispatches DRAM-ready requests without waiting behind in-flight SSD restores
in the \textbf{SSD-Preparing Queue}, orders pending SSD prefetches by transfer size and wait time via Eq.~\ref{eq:priority},
and re-admits completed prefetches using their original arrival timestamp~(\texttt{orig\_arrival\_ts})
so DRAM bursts cannot starve SSD-backed sessions.

\begin{figure}[!t]
  \centering  \includegraphics[width=\columnwidth]{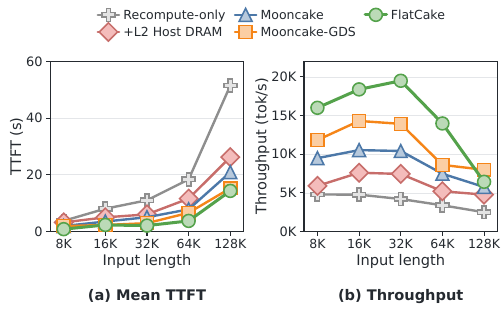}
  \vspace{-10mm}
  \caption{Performance on Llama-3.1-70B~(TP4 on 4$\times$A100 GPUs) across input context lengths from 8K to 128K tokens:
  (a)~mean TTFT and (b)~token throughput.}
  \Description{Two line charts show Llama-3.1-70B performance with TP4 on
  A100 GPUs across input lengths from 8K to 128K tokens. The left panel
  reports mean TTFT, and the right panel reports throughput. Both panels
  compare Recompute-only, GPU HBM with L2 Host DRAM, Mooncake, Mooncake-GDS,
  and FlatCake.}
  \label{fig:llama70b_sensitivity}
\end{figure}

\subsection{Cross-Model and Context-Length Scaling}
\label{sec:eval_sensitivity}
To answer \textbf{E3}, we evaluate whether \sys's layout and streaming gains generalize
from hybrid compressed attention~(DeepSeek-V4-Flash on B200) to standard grouped-query attention~(GQA),
and whether adding direct GDS transport to a tiered hierarchy is sufficient on long contexts.
Because \cio organizes KV blocks by session scope in layer-major order~(\S\ref{sec:gds}),
each layer's prefix forms a contiguous SSD extent regardless of per-layer head dimensions.
Figure~\ref{fig:llama70b_sensitivity} compares \sys against \textbf{Recompute-only}, \textbf{+L2 Host DRAM}, \textbf{Mooncake}, and \textbf{Mooncake-GDS}
on Llama-3.1-70B~(TP4 on four A100 GPUs) across input lengths from 8K to 128K tokens.
\sys reduces mean TTFT by $1.5\times$--$2.4\times$ over Mooncake and by up to $2.3\times$ over Mooncake-GDS.
Mooncake-GDS ties \sys at 16K~(2.31 vs.\ 2.32\,s) but is $1.8\times$ slower at 64K~(6.60 vs.\ 3.74\,s)
with $1.6\times$ lower throughput~(8.6K vs.\ 14.0K\,tokens/s);
at 128K the two converge~(15.3 vs.\ 14.3\,s) and Mooncake-GDS attains higher throughput~(8.0K vs.\ 6.4K\,tokens/s).
Thus GDS alone recovers much of the transport cost, but an eviction-scrambled layout
still leaves it behind \sys at mid-length contexts.

\section{Related Work}
\label{sec:related}

\rih{Tiered KV Cache Storage and Offloading}
Existing LLM serving systems extend HBM capacity by offloading KV blocks to host DRAM and NVMe SSD
through a tiered hierarchy~\cite{flexgen2023,alisa2024,flashgen2025,cachedattention2024,mooncake2024,lmcache2025,sglang_hicache2025,hifc2025},
building on general-purpose tiered memory designs~\cite{flatflash2019,g102023,deepum2023,ariadne2026,aifm2020,term2024,memtis2023,tpp2023,nomad2024,m52025,tmo2022,warehouse_tiering2023}.
Recent studies characterize production KV reuse~\cite{kvcache_wild2026}
and improve tiered hierarchies via workload-aware eviction~(Bidaw~\cite{bidaw2026}),
selective token or layer offloading~\cite{impress2025,infinigen2024,cxl_vm2024,diffkv2021},
joint compression and tier placement~(AdaptCache~\cite{adaptcache2025}),
or SSD-aware tier coordination~(SolidAttention~\cite{solidattention2026}, KVDrive~\cite{kvdrive2026}).
Despite these optimizations, all of these systems retain a migratory hierarchy in which blocks enter host DRAM first
and reach SSD through incremental eviction or demotion.
In contrast, \sys exploits the write-once, session-scoped immutability of KV blocks
to bind each block directly to DRAM or SSD at creation time~(\gva) and never migrate it thereafter.

\rih{Direct Storage I/O and Request Scheduling}
Several systems accelerate the SSD-to-GPU read path using GPUDirect Storage~\cite{spin2017,flashneuron2021,bam2023,geminifs2025,gofs2025},
pre-registered GPU buffers with parallel disk workers~(LMCache~\cite{lmcache2025}),
GPU-initiated \texttt{io\_uring} queues~(Tutti~\cite{tutti2025}),
or storage-near execution~\cite{instattention2025,directkv2026}.
However, optimizing the transport primitive alone does not resolve on-disk fragmentation:
when SSD space is populated in LRU eviction order, a session's prefix remains scattered across non-contiguous LBAs,
preventing NVMe controllers from coalescing reads into single DMA transfers~(Observation~\ref{obs:twohop}).
\sys co-designs write-time layer-major scope layout within pre-allocated \texttt{GPU File}s
and read-time sort-merge coalescing~(\cio) so that standard GDS transfers read sequential per-layer extents.
Furthermore, unlike device-level I/O schedulers~\cite{d2fq2021},
\mas uses the media identity encoded in each 64-bit GVA to decouple DRAM-ready and SSD-preparing queues
while preserving arrival-order fairness across media.

\section{Conclusion}
\label{sec:conclusion}
For write-once, prefix-grouped LLM KV caches, placing host DRAM as a migratory cache and staging layer in front of NVMe SSD
pollutes memory and fragments disk layout, whereas omitting DRAM leaves fast server memory idle.
\sys instead organizes host DRAM and NVMe SSD as peer media behind a unified address space.
By combining write-time placement~(\gva), medium-aware scheduling~(\mas), and layer-major direct streaming~(\cio),
\sys keeps compact prefixes in DRAM and streams large session histories sequentially from SSD into a bounded GPU window without cross-media migration.
Across three production traces~(DeepSeek-V4-Flash) and long-context workloads up to 128K tokens~(Llama-3.1-70B),
\sys prevents high-load hit-rate collapse, reduces mean TTFT by up to $2.4\times$,
and delivers up to $2.16\times$ higher token throughput over tiered and SSD-direct baselines.
These results show that write-time peer placement and layer-major layout, rather than transport bypass alone, are key to SSD-backed KV caching.


\balance
\bibliographystyle{ACM-Reference-Format}
\bibliography{refs}

@inproceedings{mooncake2024,
  author    = {Qin, Ruoyu and Li, Zheming and He, Weiran and Cui, Jialei and Ren, Feng and Zhang, Mingxing and Wu, Yongwei and Zheng, Weimin and Xu, Xinran},
  title     = {Mooncake: Trading More Storage for Less Computation---A {KVCache}-centric Architecture for Serving {LLM} Chatbot},
  booktitle = {Proceedings of the 23rd USENIX Conference on File and Storage Technologies (FAST)},
  publisher = {USENIX Association},
  address   = {Santa Clara, CA, USA},
  pages     = {155--170},
  year      = {2025},
  url       = {https://www.usenix.org/conference/fast25/presentation/qin}
}

@inproceedings{vllm2023,
  author    = {Kwon, Woosuk and Li, Zhuohan and Zhuang, Siyuan and Sheng, Ying and Zheng, Lianmin and Yu, Cody Hao and Gonzalez, Joseph E. and Zhang, Hao and Stoica, Ion},
  title     = {Efficient Memory Management for Large Language Model Serving with {PagedAttention}},
  booktitle = {Proceedings of the 29th Symposium on Operating Systems Principles (SOSP)},
  publisher = {ACM},
  address   = {Koblenz, Germany},
  pages     = {611--626},
  year      = {2023}
}

@inproceedings{sglang2024,
  author    = {Zheng, Lianmin and Yin, Liangsheng and Xie, Zhiqiang and Sun, Chuyue and Huang, Jeff and Yu, Cody Hao and Cao, Shiyi and Kozyrakis, Christos and Stoica, Ion and Gonzalez, Joseph E. and Barrett, Clark and Sheng, Ying},
  title     = {{SGLang}: Efficient Execution of Structured Language Model Programs},
  booktitle = {Proceedings of the Annual Conference on Neural Information Processing Systems (NeurIPS)},
  publisher = {NeurIPS},
  address   = {Vancouver, Canada},
  pages     = {62557--62583},
  year      = {2024}
}

@inproceedings{kvcache_wild2026,
  author    = {Wang, Jiahao and Han, Jinbo and Wei, Xingda and Shen, Sijie and
               Zhang, Dingyan and Fang, Chenguang and Chen, Rong and Yu, Wenyuan
               and Chen, Haibo},
  title     = {{KVCache} Cache in the Wild: Characterizing and Optimizing {KVCache} Cache at a Large Cloud Provider},
  booktitle = {Proceedings of the 2025 USENIX Annual Technical Conference (ATC)},
  publisher = {USENIX Association},
  address   = {Boston, MA},
  pages     = {465--482},
  year      = {2025}
}

@inproceedings{bidaw2026,
  author    = {Hu, Shipeng and Zhang, Guangyan and Zhou, Yuqi and Wei, Yaya and Zhong, Ziyan and Chen, Jike},
  title     = {Bidaw: Enhancing Key-Value Caching for Interactive {LLM} Serving via Bidirectional Computation--Storage Awareness},
  booktitle = {Proceedings of the 24th USENIX Conference on File and Storage Technologies (FAST)},
  publisher = {USENIX Association},
  address   = {Santa Clara, CA, USA},
  pages     = {101--116},
  year      = {2026}
}

@inproceedings{impress2025,
  author    = {Chen, Weijian and He, Shuibing and Qu, Haoyang and Zhang, Ruidong and Yang, Siling and Chen, Ping and Zheng, Yi and Huai, Baoxing and Chen, Gang},
  title     = {{IMPRESS}: An Importance-Informed Multi-Tier Prefix {KV} Storage System for Large Language Model Inference},
  booktitle = {Proceedings of the 23rd USENIX Conference on File and Storage Technologies (FAST)},
  publisher = {USENIX Association},
  address   = {Santa Clara, CA, USA},
  pages     = {187--201},
  year      = {2025},
  url       = {https://www.usenix.org/conference/fast25/presentation/chen-weijian-impress}
}

@inproceedings{iochar2025,
  author    = {Ren, Zebin and Pinto, Christian and Doekemeijer, Krijn and
               De Matteis, Tiziano and Stoica, Radu and Trivedi, Animesh},
  title     = {An {I/O} Characterizing Study of Offloading {LLM} Models and {KV} Caches to {NVMe} {SSD}},
  booktitle = {Proceedings of the 5th Workshop on Challenges and Opportunities
               of Efficient and Performant Storage Systems (CHEOPS)},
  publisher = {ACM},
  address   = {Rotterdam, Netherlands},
  pages     = {23--33},
  year      = {2025}
}

@inproceedings{gofs2025,
  author    = {Li, Shaobo and Zhou, Yirui Eric and Xue, Yuqi and Xu, Yuan and Huang, Jian},
  title     = {Managing Scalable Direct Storage Accesses for {GPUs} with {GoFS}},
  booktitle = {Proceedings of the 31st ACM SIGOPS Symposium on Operating Systems Principles (SOSP)},
  publisher = {ACM},
  address   = {Seoul, Republic of Korea},
  pages     = {979--995},
  year      = {2025}
}

@inproceedings{infinigen2024,
  author    = {Lee, Wonbeom and Lee, Jungi and Seo, Junghwan and Sim, Jaewoong},
  title     = {{InfiniGen}: Efficient Generative Inference of Large Language Models with Dynamic {KV} Cache Management},
  booktitle = {Proceedings of the 18th USENIX Symposium on Operating Systems Design and Implementation (OSDI)},
  publisher = {USENIX Association},
  address   = {Santa Clara, CA, USA},
  pages     = {155--172},
  year      = {2024},
  url       = {https://www.usenix.org/conference/osdi24/presentation/lee}
}

@inproceedings{diffkv2021,
  author    = {Li, Yongkun and Liu, Zhen and Lee, Patrick P. C. and Wu, Jiayu and Xu, Yinlong and Wu, Yi and Tang, Liu and Liu, Qi and Cui, Qiu},
  title     = {Differentiated {Key-Value} Storage Management for Balanced {I/O} Performance},
  booktitle = {Proceedings of the 2021 USENIX Annual Technical Conference (ATC)},
  publisher = {USENIX Association},
  address   = {Virtual},
  pages     = {673--687},
  year      = {2021},
  url       = {https://www.usenix.org/conference/atc21/presentation/li-yongkun}
}

@inproceedings{cxl_vm2024,
  author    = {Zhong, Yuhong and Berger, Daniel S. and Waldspurger, Carl and Wee, Ryan and Agarwal, Ishwar and Agarwal, Rajat and Hady, Frank and Kumar, Karthik and Hill, Mark D. and Chowdhury, Mosharaf and Cidon, Asaf},
  title     = {Managing Memory Tiers with {CXL} in Virtualized Environments},
  booktitle = {Proceedings of the 18th USENIX Symposium on Operating Systems Design and Implementation (OSDI)},
  publisher = {USENIX Association},
  address   = {Santa Clara, CA, USA},
  pages     = {37--56},
  year      = {2024},
  url       = {https://www.usenix.org/conference/osdi24/presentation/zhong-yuhong}
}

@inproceedings{attention2017,
  author    = {Vaswani, Ashish and Shazeer, Noam and Parmar, Niki and Uszkoreit, Jakob
               and Jones, Llion and Gomez, Aidan N. and Kaiser, {\L}ukasz
               and Polosukhin, Illia},
  title     = {Attention Is All You Need},
  booktitle = {Advances in Neural Information Processing Systems 30 ({NIPS} 2017)},
  pages     = {5998--6008},
  address   = {Long Beach, CA, USA},
  year      = {2017},
  publisher = {Curran Associates, Inc.}
}

@misc{lmcache2025,
  author  = {Liu, Yuhan and Cheng, Yihua and Yao, Jiayi and An, Yuwei and Chen, Xiaokun and Feng, Shaoting and Huang, Yuyang and Shen, Samuel and Zhang, Rui and Du, Kuntai and Jiang, Junchen},
  title   = {{LMCache}: An Efficient {KV} Cache Layer for Enterprise-Scale {LLM} Inference},
  eprint       = {2510.09665},
  archivePrefix= {arXiv},
  eprinttype   = {arXiv},
  year    = {2025},
  howpublished = {arXiv preprint}
}

@inproceedings{flexgen2023,
  author    = {Sheng, Ying and Zheng, Lianmin and Yuan, Binhang and Li, Zhuohan and Ryabinin, Max and Chen, Beidi and Liang, Percy and R{\'e}, Christopher and Stoica, Ion and Zhang, Ce},
  title     = {{FlexGen}: High-Throughput Generative Inference of Large Language Models with a Single {GPU}},
  booktitle = {Proceedings of the 40th International Conference on Machine Learning (ICML)},
  publisher = {PMLR},
  pages     = {31094--31116},
  address   = {Honolulu, HI, USA},
  year      = {2023}
}

@misc{tutti2025,
  author       = {Shi Qiu and Yifan Hu and Xintao Wang and Wenhao Zhu and Jianqin Yan and Hao Chen and Kaiqiang Xu and Kai Chen and Yiming Zhang},
  title        = {Tutti: Making {SSD}-Backed {KV} Cache Practical for Long-Context {LLM} Serving},
  eprint       = {2605.03375},
  archivePrefix= {arXiv},
  eprinttype   = {arXiv},
  year         = {2025},
  howpublished = {arXiv preprint}
}

@misc{sglang_hicache2025,
  author       = {Zhiqiang Xie and {SGLang Team}},
  title        = {{SGLang} {HiCache}: Fast Hierarchical {KV} Caching with Your Favorite Storage Backends},
  howpublished = {\url{https://lmsys.org/blog/2025-09-10-sglang-hicache/}},
  year         = {2025}
}

@misc{ccTracesWeka2026,
  author       = {{SemiAnalysis}},
  title        = {{CC Traces --- Weka, With Subagents (v7)}},
  year         = {2026},
  howpublished = {\url{https://huggingface.co/datasets/semianalysisai/cc-traces-weka-062126}},
  note         = {Claude Code agentic coding traces; accessed 2026-07-28}
}

@misc{deepseekv4,
  author  = {{DeepSeek-AI}},
  title   = {{DeepSeek-V4: Towards Highly Efficient Million-Token Context Intelligence}},
  eprint       = {2606.19348},
  archivePrefix= {arXiv},
  eprinttype   = {arXiv},
  year    = {2026},
  note    = {\S3.5.2 ``On-Disk KV Cache Storage''},
  howpublished = {arXiv preprint}
}

@inproceedings{g102023,
  author    = {Zhang, Haoyang and Zhou, Yirui Eric and Xue, Yuqi and Liu, Yiqi and Huang, Jian},
  title     = {{G10}: Enabling an Efficient Unified {GPU} Memory and Storage Architecture with Smart Tensor Migrations},
  booktitle = {Proceedings of the 56th Annual {IEEE/ACM} International Symposium on Microarchitecture},
  pages     = {395--410},
  address   = {Toronto, ON, Canada},
  year      = {2023},
  publisher = {Association for Computing Machinery},
  doi       = {10.1145/3613424.3614309},
  url       = {https://doi.org/10.1145/3613424.3614309}
}

@inproceedings{flatflash2019,
  author    = {Abulila, Ahmed H. M. O. and Mailthody, Vikram Sharma and Qureshi, Zaid and Huang, Jian and Kim, Nam Sung and Xiong, Jinjun and Hwu, Wen-Mei W.},
  title     = {{FlatFlash}: Exploiting the Byte-Accessibility of {SSDs} within a Unified Memory-Storage Hierarchy},
  booktitle = {Proceedings of the 24th International Conference on Architectural Support for Programming Languages and Operating Systems},
  pages     = {971--985},
  address   = {Providence, RI, USA},
  year      = {2019},
  publisher = {Association for Computing Machinery},
  doi       = {10.1145/3297858.3304061},
  url       = {https://doi.org/10.1145/3297858.3304061}
}

@inproceedings{flashgen2025,
  author    = {Jeong, Jinwoo and Ahn, Jeongseob},
  title     = {Accelerating {LLM} Serving for Multi-turn Dialogues with Efficient Resource Management},
  booktitle = {Proceedings of the 30th {ACM} International Conference on Architectural Support for Programming Languages and Operating Systems, Volume 2},
  pages     = {1--15},
  address   = {Rotterdam, Netherlands},
  year      = {2025},
  publisher = {Association for Computing Machinery},
  doi       = {10.1145/3676641.3716245},
  url       = {https://doi.org/10.1145/3676641.3716245}
}

@inproceedings{cachedattention2024,
  author    = {Gao, Bin and He, Zhuomin and Sharma, Puru and Kang, Qingxuan and Jevdjic, Djordje and Deng, Junbo and Yang, Xingkun and Yu, Zhou and Zuo, Pengfei},
  title     = {Cost-Efficient Large Language Model Serving for Multi-turn Conversations with {CachedAttention}},
  booktitle = {2024 USENIX Annual Technical Conference (USENIX ATC 24)},
  pages     = {111--126},
  address   = {Santa Clara, CA, USA},
  year      = {2024},
  publisher = {USENIX Association},
  url       = {https://www.usenix.org/conference/atc24/presentation/gao-bin-cost}
}

@inproceedings{directkv2026,
  author    = {Luo, Shutian and Shen, Haiying},
  title     = {No Buffer, No Bottleneck: Efficient Zero-Copy {KV} Cache Offloading for Long-Context {LLMs}},
  booktitle = {20th USENIX Symposium on Operating Systems Design and Implementation (OSDI 26)},
  pages     = {39--54},
  address   = {Seattle, WA, USA},
  year      = {2026},
  publisher = {USENIX Association},
  url       = {https://www.usenix.org/conference/osdi26/presentation/luo}
}

@inproceedings{alisa2024,
  author    = {Zhao, Youpeng and Wu, Di and Wang, Jun},
  title     = {{ALISA}: Accelerating Large Language Model Inference via Sparsity-Aware {KV} Caching},
  booktitle = {2024 {ACM/IEEE} 51st Annual International Symposium on Computer Architecture (ISCA)},
  pages     = {1005--1017},
  address   = {Buenos Aires, Argentina},
  year      = {2024},
  publisher = {IEEE},
  doi       = {10.1109/ISCA59077.2024.00077},
  url       = {https://doi.org/10.1109/ISCA59077.2024.00077}
}

@inproceedings{instattention2025,
  author    = {Pan, Xiurui and Li, Endian and Li, Qiao and Liang, Shengwen and Shan, Yizhou and Zhou, Ke and Luo, Yingwei and Wang, Xiaolin and Zhang, Jie},
  title     = {{InstAttention}: In-Storage Attention Offloading for Cost-Effective Long-Context {LLM} Inference},
  booktitle = {2025 {IEEE} International Symposium on High Performance Computer Architecture (HPCA)},
  pages     = {1510--1525},
  address   = {Las Vegas, NV, USA},
  year      = {2025},
  publisher = {IEEE},
  doi       = {10.1109/HPCA61900.2025.00113},
  url       = {https://doi.org/10.1109/HPCA61900.2025.00113}
}

@inproceedings{bam2023,
  author    = {Qureshi, Zaid and Mailthody, Vikram Sharma and Gelado, Isaac and Min, Seung Won and Masood, Amna and Park, Jeongmin and Xiong, Jinjun and Newburn, C. J. and Vainbrand, Dmitri and Chung, I-Hsin and Garland, Michael and Dally, William J. and Hwu, Wen-Mei W.},
  title     = {{GPU}-Initiated On-Demand High-Throughput Storage Access in the {BaM} System Architecture},
  booktitle = {Proceedings of the 28th {ACM} International Conference on Architectural Support for Programming Languages and Operating Systems, Volume 2},
  pages     = {325--339},
  address   = {Vancouver, BC, Canada},
  year      = {2023},
  publisher = {Association for Computing Machinery},
  doi       = {10.1145/3575693.3575748},
  url       = {https://doi.org/10.1145/3575693.3575748}
}

@inproceedings{geminifs2025,
  author    = {Qiu, Shi and Liu, Weinan and Hu, Yifan and Yan, Jianqin and Shen, Zhirong and Yao, Xin and Chen, Renhai and Zhang, Gong and Zhang, Yiming},
  title     = {{GeminiFS}: A Companion File System for {GPUs}},
  booktitle = {23rd USENIX Conference on File and Storage Technologies (FAST 25)},
  pages     = {221--236},
  address   = {Santa Clara, CA, USA},
  year      = {2025},
  publisher = {USENIX Association},
  url       = {https://www.usenix.org/conference/fast25/presentation/qiu}
}

@inproceedings{nomad2024,
  author    = {Xiang, Lingfeng and Lin, Zhen and Deng, Weishu and Lu, Hui and Rao, Jia and Yuan, Yifan and Wang, Ren},
  title     = {{NOMAD}: Non-Exclusive Memory Tiering via Transactional Page Migration},
  booktitle = {18th USENIX Symposium on Operating Systems Design and Implementation (OSDI 24)},
  pages     = {19--35},
  address   = {Santa Clara, CA, USA},
  year      = {2024},
  publisher = {USENIX Association},
  url       = {https://www.usenix.org/conference/osdi24/presentation/xiang}
}

@inproceedings{memtis2023,
  author    = {Lee, Taehyung and Monga, Sumit Kumar and Min, Changwoo and Eom, Young Ik},
  title     = {{MEMTIS}: Efficient Memory Tiering with Dynamic Page Classification and Page Size Determination},
  booktitle = {Proceedings of the 29th Symposium on Operating Systems Principles},
  pages     = {17--34},
  address   = {Koblenz, Germany},
  year      = {2023},
  publisher = {Association for Computing Machinery},
  doi       = {10.1145/3600006.3613167},
  url       = {https://doi.org/10.1145/3600006.3613167}
}

@inproceedings{tpp2023,
  author    = {Al Maruf, Hasan and Wang, Hao and Dhanotia, Abhishek and Weiner, Johannes and Agarwal, Niket and Bhattacharya, Pallab and Petersen, Chris and Chowdhury, Mosharaf and Kanaujia, Shobhit O. and Chauhan, Prakash},
  title     = {{TPP}: Transparent Page Placement for {CXL}-Enabled Tiered-Memory},
  booktitle = {Proceedings of the 28th {ACM} International Conference on Architectural Support for Programming Languages and Operating Systems, Volume 3},
  pages     = {742--755},
  address   = {Vancouver, BC, Canada},
  year      = {2023},
  publisher = {Association for Computing Machinery},
  doi       = {10.1145/3582016.3582063},
  url       = {https://doi.org/10.1145/3582016.3582063}
}

@inproceedings{m52025,
  author    = {Sun, Yan and Kim, Jongyul and Yu, Zeduo and Zhang, Jiyuan and Chai, Siyuan and Kim, Michael Jaemin and Nam, Hwayong and Park, Jaehyun and Na, Eojin and Yuan, Yifan and Wang, Ren and Ahn, Jung Ho and Xu, Tianyin and Kim, Nam Sung},
  title     = {{M5}: Mastering Page Migration and Memory Management for {CXL}-based Tiered Memory Systems},
  booktitle = {Proceedings of the 30th {ACM} International Conference on Architectural Support for Programming Languages and Operating Systems, Volume 2},
  pages     = {604--621},
  address   = {Rotterdam, Netherlands},
  year      = {2025},
  publisher = {Association for Computing Machinery},
  doi       = {10.1145/3676641.3711999},
  url       = {https://doi.org/10.1145/3676641.3711999}
}

@inproceedings{tmo2022,
  author    = {Weiner, Johannes and Agarwal, Niket and Schatzberg, Dan and Yang, Leon and Wang, Hao and Sanouillet, Blaise and Sharma, Bikash and Heo, Tejun and Jain, Mayank and Tang, Chunqiang and Skarlatos, Dimitrios},
  title     = {{TMO}: Transparent Memory Offloading in Datacenters},
  booktitle = {Proceedings of the 27th {ACM} International Conference on Architectural Support for Programming Languages and Operating Systems},
  pages     = {609--621},
  address   = {Lausanne, Switzerland},
  year      = {2022},
  publisher = {Association for Computing Machinery},
  doi       = {10.1145/3503222.3507731},
  url       = {https://doi.org/10.1145/3503222.3507731}
}

@inproceedings{deepum2023,
  author    = {Jung, Jaehoon and Kim, Jinpyo and Lee, Jaejin},
  title     = {{DeepUM}: Tensor Migration and Prefetching in Unified Memory},
  booktitle = {Proceedings of the 28th {ACM} International Conference on Architectural Support for Programming Languages and Operating Systems, Volume 2},
  pages     = {207--221},
  address   = {Vancouver, BC, Canada},
  year      = {2023},
  publisher = {Association for Computing Machinery},
  doi       = {10.1145/3575693.3575736},
  url       = {https://doi.org/10.1145/3575693.3575736}
}

@inproceedings{ariadne2026,
  author    = {Shin, Hyunkyun and Bang, Seongtae and Park, Hyungwon and Kim, Daehoon},
  title     = {{ARIADNE}: Adaptive {UVM} Management for Efficient {GPU} Memory Oversubscription},
  booktitle = {2026 {IEEE} International Symposium on High Performance Computer Architecture (HPCA)},
  pages     = {1--15},
  address   = {Sydney, Australia},
  year      = {2026},
  publisher = {IEEE},
  doi       = {10.1109/HPCA68181.2026.11408564},
  url       = {https://doi.org/10.1109/HPCA68181.2026.11408564}
}

@inproceedings{warehouse_tiering2023,
  author    = {Duraisamy, Padmapriya and Xu, Wei and Hare, Scott and Rajwar, Ravi and Culler, David E. and Xu, Zhiyi and Fan, Jianing and Kennelly, Christopher and McCloskey, Bill and Mijailovic, Danijela and Morris, Brian and Mukherjee, Chiranjit and Ren, Jingliang and Thelen, Greg and Turner, Paul and Villavieja, Carlos and Ranganathan, Parthasarathy and Vahdat, Amin},
  title     = {Towards an Adaptable Systems Architecture for Memory Tiering at Warehouse-Scale},
  booktitle = {Proceedings of the 28th {ACM} International Conference on Architectural Support for Programming Languages and Operating Systems, Volume 3},
  pages     = {727--741},
  address   = {Vancouver, BC, Canada},
  year      = {2023},
  publisher = {Association for Computing Machinery},
  doi       = {10.1145/3582016.3582031},
  url       = {https://doi.org/10.1145/3582016.3582031}
}

@inproceedings{aifm2020,
  author    = {Ruan, Zhenyuan and Schwarzkopf, Malte and Aguilera, Marcos K. and Belay, Adam},
  title     = {{AIFM}: High-Performance, Application-Integrated Far Memory},
  booktitle = {14th USENIX Symposium on Operating Systems Design and Implementation (OSDI 20)},
  pages     = {315--332},
  address   = {Virtual},
  year      = {2020},
  publisher = {USENIX Association},
  url       = {https://www.usenix.org/conference/osdi20/presentation/ruan}
}

@inproceedings{term2024,
  author    = {Yang, Zhe and Wang, Qing and Liao, Xiaojian and Lu, Youyou and Huang, Keji and Shu, Jiwu},
  title     = {{TeRM}: Extending {RDMA}-Attached Memory with {SSD}},
  booktitle = {22nd USENIX Conference on File and Storage Technologies (FAST 24)},
  pages     = {1--16},
  address   = {Santa Clara, CA, USA},
  year      = {2024},
  publisher = {USENIX Association},
  url       = {https://www.usenix.org/conference/fast24/presentation/yang-zhe}
}

@inproceedings{spin2017,
  author    = {Bergman, Shai and Brokhman, Tanya and Cohen, Tzachi and Silberstein, Mark},
  title     = {{SPIN}: Seamless Operating System Integration of Peer-to-Peer {DMA} Between {SSDs} and {GPUs}},
  booktitle = {2017 USENIX Annual Technical Conference (USENIX ATC 17)},
  pages     = {167--179},
  address   = {Santa Clara, CA, USA},
  year      = {2017},
  publisher = {USENIX Association},
  url       = {https://www.usenix.org/conference/atc17/technical-sessions/presentation/bergman}
}

@inproceedings{flashneuron2021,
  author    = {Bae, Jonghyun and Lee, Jongsung and Jin, Yunho and Son, Sam and Kim, Shine and Jang, Hakbeom and Ham, Tae Jun and Lee, Jae W.},
  title     = {{FlashNeuron}: {SSD}-Enabled Large-Batch Training of Very Deep Neural Networks},
  booktitle = {19th USENIX Conference on File and Storage Technologies (FAST 21)},
  pages     = {387--401},
  address   = {Santa Clara, CA, USA},
  year      = {2021},
  publisher = {USENIX Association},
  url       = {https://www.usenix.org/conference/fast21/presentation/bae}
}

@inproceedings{d2fq2021,
  author    = {Woo, Jiwon and Ahn, Minwoo and Lee, Gyusun and Jeong, Jinkyu},
  title     = {{D2FQ}: Device-Direct Fair Queueing for {NVMe} {SSDs}},
  booktitle = {19th USENIX Conference on File and Storage Technologies (FAST 21)},
  pages     = {403--415},
  address   = {Santa Clara, CA, USA},
  year      = {2021},
  publisher = {USENIX Association},
  url       = {https://www.usenix.org/conference/fast21/presentation/woo}
}

@inproceedings{hrrn1971,
  author    = {Brinch Hansen, Per},
  title     = {Short-term Scheduling in Multiprogramming Systems},
  booktitle = {Proceedings of the 3rd {ACM} Symposium on Operating Systems
               Principles ({SOSP} '71)},
  pages     = {103--105},
  address   = {Palo Alto, CA, USA},
  year      = {1971},
  publisher = {Association for Computing Machinery}
}

@inproceedings{hifc2025,
  author    = {Jeong, Inho and Woo, Sunghyeon and Namkung, Sol and Jeon, Dongsuk},
  title     = {{HiFC}: High-efficiency Flash-based {KV} Cache Swapping for Scaling {LLM} Inference},
  booktitle = {Advances in Neural Information Processing Systems},
  volume    = {38},
  pages     = {47561--47590},
  address   = {San Diego, CA, USA},
  year      = {2025},
  publisher = {Curran Associates, Inc.},
  doi       = {10.52202/085713-1587},
  url       = {https://proceedings.neurips.cc/paper_files/paper/2025/hash/4431224d3762aa655f0aee4eaf04ff16-Abstract-Conference.html}
}

@misc{adaptcache2025,
  author       = {Feng, Shaoting and Li, Hanchen and Du, Kuntai and Gu, Zhuohan and Liu, Yuhan and Yao, Jiayi and Ray, Siddhant and Shen, Samuel and Cheng, Yihua and Ananthanarayanan, Ganesh and Jiang, Junchen},
  title        = {{AdaptCache}: {KV} Cache Native Storage Hierarchy for Low-Delay and High-Quality Language Model Serving},
  howpublished = {ACM SOSP BigMem Workshop},
  year         = {2025},
  note         = {Non-archival workshop paper},
  url          = {https://www.microsoft.com/en-us/research/publication/adaptcache-kv-cache-native-storage-hierarchy-for-low-delay-and-high-quality-language-model-serving/}
}

@inproceedings{solidattention2026,
  author    = {Zheng, Xinrui and Wei, Dongliang and Gao, Jianxiang and Song, Yixin and Mi, Zeyu and Chen, Haibo},
  title     = {{SolidAttention}: Low-Latency {SSD}-based Serving on Memory-Constrained {PCs}},
  booktitle = {24th USENIX Conference on File and Storage Technologies (FAST 26)},
  pages     = {67--82},
  address   = {Santa Clara, CA, USA},
  year      = {2026},
  publisher = {USENIX Association},
  url       = {https://www.usenix.org/conference/fast26/presentation/zheng}
}

@misc{kvdrive2026,
  author       = {Lin, Jian and Mi, Jiazhi and Hong, Zicong and Wang, Haodong and Liu, Qianli and Zhang, Haodyue and Li, Peng and Guo, Song},
  title        = {{KVDrive}: A Holistic Multi-Tier {KV} Cache Management System for Long-Context {LLM} Inference},
  eprint       = {2605.18071},
  archivePrefix= {arXiv},
  eprinttype   = {arXiv},
  year         = {2026},
  howpublished = {arXiv preprint},
  doi          = {10.48550/arXiv.2605.18071},
  url          = {https://arxiv.org/abs/2605.18071}
}

\end{document}